\documentclass[twocolumn]{aastex701}
\usepackage{siunitx}

\usepackage{booktabs}

\begin{document}

\title{Potassium abundances in extremely metal poor stars: Implications for nucleosynthesis in the final stages of massive star evolution\footnote{This research is based on data collected at the Subaru Telescope, which is operated by the National Astronomical Observatory of Japan. We are honored and grateful for the opportunity of observing the Universe from Maunakea, which has the cultural, historical, and natural significance in Hawaii.}}

\author[0000-0003-4656-0241,sname=Ishigaki,gname=Miho]{Miho N. Ishigaki}
\affiliation{Subaru Telescope, National Astronomical Observatory of Japan \\
2-21-1, Mitaka, Tokyo 181-8588, Japan}
\affiliation{Kavli Institute for the Physics and Mathematics of the Universe (WPI), The University of Tokyo Institutes for Advanced Study, The University of Tokyo, Kashiwa, Chiba 277-8583, Japan}
\affiliation{Department of Astronomical Science, SOKENDAI (The Graduate University for Advanced Studies), 2-21-1 Osawa, Mitaka, Tokyo, 181-8588, Japan}
\email{miho.ishigaki@nao.ac.jp}

\author[0000-0001-8537-3153]{Nozomu Tominaga}
\affiliation{Division of Science, National Astronomical Observatory of Japan \\
2-21-1, Mitaka, Tokyo 181-8588, Japan}
\affiliation{Kavli Institute for the Physics and Mathematics of the Universe (WPI), The University of Tokyo Institutes for Advanced Study, The University of Tokyo, Kashiwa,
Chiba 277-8583, Japan}
\affiliation{Department of Astronomical Science, SOKENDAI (The Graduate University for Advanced Studies), 2-21-1 Osawa, Mitaka, Tokyo, 181-8588, Japan}
\email{nozomu.tominaga@nao.ac.jp}

\author[0000-0002-8975-6829]{Wako Aoki}
\affiliation{National Astronomical Observatory of Japan \\
2-21-1, Mitaka, Tokyo 181-8588, Japan}
\affiliation{Department of Astronomical Science, SOKENDAI (The Graduate University for Advanced Studies), 2-21-1 Osawa, Mitaka, Tokyo, 181-8588, Japan}
\email{aoki.wako@nao.ac.jp}

\author[0000-0002-4759-7794]{Shinya Wanajo}
\affiliation{Astronomical Institute, Graduate School of Science, Tohoku University, \\
6-3 Aoba, Aramaki, Aoba-ku, Sendai, Miyagi 980-8578, Japan}
\email{shinya.wanajo@astr.tohoku.ac.jp}

\author[0000-0003-0304-9283]{Tomoya Takiwaki}
\affiliation{National Astronomical Observatory of Japan \\
2-21-1, Mitaka, Tokyo 181-8588, Japan}
\affiliation{Department of Astronomical Science, SOKENDAI (The Graduate University for Advanced Studies), 2-21-1 Osawa, Mitaka, Tokyo, 181-8588, Japan}
\email{takiwaki.tomoya.astro@gmail.com}

\author[0000-0002-8734-2147]{Ko Nakamura}
\affiliation{Department of Applied Physics, Fukuoka University, \\
Nanakuma Jonan 8-19-1, Fukuoka 814-0180, Japan}
\affiliation{Research Institute of Stellar Explosive Phenomena, Fukuoka University, \\Nanakuma Jonan 8-19-1, Fukuoka 814-0180, Japan}
\email{nakamurako@fukuoka-u.ac.jp}

\author{Nobuyuki Iwamoto}
\affiliation{Nuclear Science and Engineering Center, Japan Atomic Energy Agency, \\
2-4 Shirakata, Tokai, Naka, Ibaraki 319-1195, Japan}
\email{iwamoto.nobuyuki@jaea.go.jp}

\author[0000-0001-9553-0685]{Ken'ichi Nomoto}
\affiliation{Kavli Institute for the Physics and Mathematics of the Universe (WPI), The University of Tokyo Institutes for Advanced Study, The University of Tokyo, Kashiwa,
Chiba 277-8583, Japan}
\email{nomoto@astron.s.u-tokyo.ac.jp}

\author[0000-0002-4343-0487]{Chiaki Kobayashi}
\affiliation{Centre for Astrophysics Research, Department of Physics, Astronomy and Mathematics, University of Hertfordshire, Hat eld, AL10 9AB, UK}
\affiliation{Kavli Institute for the Physics and Mathematics of the Universe (WPI), The University of Tokyo Institutes for Advanced Study, The University of Tokyo, Kashiwa, Chiba 277-8583, Japan}
\email{c.kobayashi@herts.ac.uk}




\begin{abstract}

We present a potassium (K) abundance analysis in extremely metal-poor (EMP) stars based on high-resolution ($R\sim$ \num{60000}) spectra obtained with the High Dispersion Spectrograph on the Subaru Telescope, covering the \ion{K}{1} resonance lines at \qty{766} and \qty{769}{\nano\meter}. One-dimensional local thermodynamic equilibrium (LTE) abundances of K and other elements, including Na, Mg, Ca, Ti, Cr, and Ni, were derived using spectral synthesis. Non-local thermodynamic equilibrium (NLTE) corrections were applied to the K abundances by interpolating a precomputed grid of corrections based on stellar parameters and the LTE K abundance. We detected \ion{K}{1} lines in seven stars with [Fe/H]~$< -3.0$ and derived upper limits for other stars in the same metallicity regime, making this sample well-suited for 
investigating the nucleosynthesis origins of K in the early universe. 
We found that the [K/Fe] and [K/Ca] ratios of the seven stars are enhanced relative to the solar value, with a scatter of approximately \num{0.1} dex, as small as the typical measurement uncertainty. 
Under the assumption
that each star formed from gas purely enriched by a single or a few 
massive stars' supernovae,  
the small scatter in [K/Fe] and [K/Ca], contrasted with the $\sim$0.7 dex scatter in [Na/Mg] ratios (after NLTE correction), suggests that the production of K in massive stars or their supernovae is independent of the processes that drive Na/Mg variation.
These findings demonstrate that K abundances in EMP stars, and their correlations with other elemental abundances, can serve as sensitive tracers of the physical mechanisms governing the final evolutionary stages of massive stars and their supernova explosions.


\end{abstract}



\section{Introduction} \label{sec:intro}

The origins of odd-atomic-numbered elements such as potassium (K, $Z=19$), scandium (Sc, $Z=21$), vanadium (V, $Z=23$) and manganese (Mn, $Z=25$) have been a topic of debate for many years, but no consensus has yet been obtained. The astrophysical origins of these elements are important not only because they are an essential ingredient for life on Earth, but also because they could serve as a tracer of extremely high temperature and density, only realized in the deep cores of massive stars or at a brief moment of supernova explosions, which are impossible to directly observe.

Potassium (K) is a particularly important element because its yield is highly sensitive to various physical processes associated with stellar and supernova nucleosynthesis. The dominant isotope of K, $^{\rm 39}$K, is predicted to be synthesized by oxygen burning in the innermost regions of massive stars \citep[e.g.,][]{Woosley_1995ApJS..101..181W}. However, Galactic 
chemical evolution models that incorporate the K yields from one-dimensional nucleosynthesis calculations of the evolution and explosion of massive stars underproduce K abundances observed in stars by more than $>1$ dex \citep[e.g.,][]{Timmes_1995ApJS...98..617T, Takeda_2009PASJ...61..563T,Andrievsky_2010A&A...509A..88A,
Nomoto_2013ARA&A..51..457N,  Kobayashi_2020ApJ...900..179K,Reggiani_2019A&A...627A.177R}. 
 Furthermore, in the old globular cluster NGC 2419, the most luminous globular cluster in the Milky Way's outer halo, an anti-correlation between K and Mg abundances has been reported \citep{Cohen2011ApJ...740...60C,Mucciarelli2012MNRAS.426.2889M}. The nucleosynthetic mechanism responsible for the synthesis of K that could account for this trend remains unclear.

Multidimensional effects in pre-supernova stellar evolution in massive stars have been suggested to enhance the production of odd Z elements, including K.  \citet{Ritter_2018MNRAS.474L...1R} used 3D hydrodynamic simulations to explore the nucleosynthesis resulting from the injection of C into the convective O-shell in the late evolution phase of massive stars with $M=25M_\odot$. 
Such C-O shell mergers are shown to explain the observed abundances of odd-Z elements in Galactic metal-poor stars.
The occurrence of shell mergers in the final stage of massive stellar evolution is further supported by recent X-ray spectroscopy of young supernova remnants, which reveals spatial variations in elemental abundance ratios predicted by this mechanism \citep[][]{Sato2025arXiv250707563S}.

Explosive nucleosynthesis that involves neutrino transport in core collapse supernovae has also been proposed to be a possible production site of odd-Z elements, including K \cite[e.g.,][]{Yoshida_2008ApJ...672.1043Y, Kobayashi_2011}. 
Explosive nucleosynthesis yields of odd-Z elements such as K, Sc, V, and Mn are generally sensitive
to the proton-to-nucleon ratio, denoted as $Y_e$, in the innermost layers of the SN ejecta \citep{Iwamoto_2006AIPC..847..409I}. The distribution of $Y_e$ in the inner-most layers of SN ejecta is determined by the interaction of neutrinos with neutrons and protons, which ultimately affects the nucleosynthesis yields. This process also determines 
the heating of matter through neutrino transport, which helps facilitate a successful explosion. Indeed, while one-dimensional simulations
of core-collapse SNe fail to reach successful explosions
\citep{Sumiyoshi_2005ApJ...629..922S}, multidimensional SN simulations indicate that a key to successful explosions is the mechanism involving those neutrino-heating processes \citep{Takiwaki_2014ApJ...786...83T,Wanajo_2011ApJ...726L..15W,Kotake_2012PTEP.2012aA301K,Bollig_2021ApJ...915...28B, Vartanyan_2024arXiv241103434V, Nakamura_2025MNRAS.536..280N}. 
Nucleosynthesis yields of core collapse supernovae based on multidimensional simulations with self-consistent treatment of neutrino transport confirmed that some of the odd-Z element isotopes
are enhanced under the high $Y_e$ environment realized in the simulation, which alleviates the discrepancy between the observed and predicted abundances of odd-Z elements \citep{Wanajo_2018ApJ...852...40W,Sieverding_2023,Wang_2024ApJ...974...39W}.

Extremely metal-poor (EMP) stars provide a unique opportunity to independently test the nucleosynthesis yields of massive stars, since their surface composition is likely determined by only a few supernovae of the first generation of massive stars \citep[e.g.][]{Audouze_1995ApJ...451L..49A}.  The chemical abundance patterns of these stars, therefore, are
an important probe of the late evolution of the cores of progenitor stars
or the physical condition (e.g. temperature, density) of stellar and supernova nucleosynthesis \citep{Umead_2005ApJ...619..427U,Tominaga_2007ApJ...660..516T,Heger_2010ApJ...724..341H,Takahashi_2014ApJ...794...40T,Tominaga_2014ApJ...785...98T,Placco_2015ApJ...809..136P,Ishigaki_2018ApJ...857...46I}.

The main reasons for the scarcity of the K abundance data for EMP stars is that the \ion{K}{1} resonance lines at \qty{766.49} and \qty{769.99}{\nano\meter}, which are the only useful K lines in optical wavelengths, are very weak in EMP stars (typical
equivalent widths of $\sim $\qty{10} - \qty{30}{\milli\angstrom}) and thus observations with
high signal-to-noise ratios are required. In practice, such a sensitivity is feasible only with \qty{8}-\qty{10}{\meter} class telescopes for the majority of EMP stars. 
Furthermore, these lines frequently overlap with strong telluric
absorption lines (e.g. O$_2$). The precise K abundance estimates in EMP stars, therefore, can only be achieved with a careful correction for the telluric absorption features.
 On top of that, the abundance estimates using the \ion{K}{1} lines are known to be largely affected by non-thermodynamic equilibrium (NLTE) effects.
Over the past decade, the number of known EMP stars
has been continuously increasing, resulting in a sample of more than $> 50$
stars with [Fe/H]$<-3.5$. \citep[e.g.,][]{Yong_2013ApJ...762...26Y, Cohen_2013ApJ...778...56C, Aoki_2013AJ....145...13A, Roderer_2014AJ....147..136R, Li_2022ApJ...931..147L}.
In these studies, however, K abundances are only sparsely reported for EMP stars because of the difficulties mentioned above.

In this paper, we present a new analysis of high-resolution spectra of \ion{K}{1} resonance lines of stars with [Fe/H]$<-3.0$. We derived [K/H] abundance ratios under the local thermodynamic equilibrium (LTE) assumption and then applied the NLTE correction on the basis of a published grid of corrections. 
We obtain an upper limit based on the observed spectra for stars whose \ion{K}{1} lines are undetectable. 

This paper is organized as follows. Section \ref{sec:obs} describes sample selection and high-resolution spectroscopic observations. Section \ref{sec:abundance_analysis} provides details of the derivation of K abundances and their NLTE correction. Analysis of other elemental abundances is also described.  Section \ref{sec:result} presents the resulting abundances and upper limits of K for the sample stars as well as their ratios with other elements. Section \ref{sec:discussion} compares observed abundances with theoretical yield models.  Finally, the conclusion is given in Section \ref{sec:conclusion}.

\section{Data} \label{sec:obs}

\subsection{Observation}
The sample stars were selected based on the studies by  \citet{Cohen_2013ApJ...778...56C}, \citet{Yong_2013ApJ...762...26Y}, \citet{Roderer_2014AJ....147..136R}, \citet{Jacobson_2015}, and \citet{Fernandez-Alvar2016A&A...593A..28F}. These studies conducted high-resolution spectroscopic surveys of a large sample of EMP stars. Stars with $V$-band magnitude $<15$ were chosen as primary targets to ensure a high signal-to-noise ratio near the K absorption lines at \qty{766.49}{\nano\meter} and \qty{769.896}{\nano\meter}. 

Observations were carried out on November 12 and 13 2016, and January 18 2017, with a high-dispersion spectrograph on the Subaru Telescope \citep{Noguchi_2002PASJ...54..855N}. The standard setting``StdRa" was used with a slit width of \ang{;;0.6}.  This setting yields the wavelength coverage of \qty{514}-\qty{637}{\nano\meter} for the blue CCD and \num{657}-\qty{778}{\nano\meter} for the red CCD with a spectral resolution of $R\sim$\num{60000}. An on-chip binning of $2\times 2$ (spatial $\times$ wavelength directions) was applied to maximize signal-to-noise ratios without degrading the spectral resolution. 
In order to identify telluric absorption lines contaminating the targets' spectra, we observed telluric standard stars frequently during the observing nights.  
Table \ref{tab:obs_info} summarizes the coordinates, date of observation, total exposure time, and magnitudes of the target stars. Signal-to-noise ratios per pixel estimated in the blue CCD and in the vicinity of the \ion{K}{1} \qty{766} and \qty{769}{\nano\meter} lines are also presented.

Radial velocities were measured by cross-correlating the observed spectra of the blue CCD with the template spectra. The spectra taken by the red CCD were not used to estimate radial velocities since the number of detectable metal absorption lines for the wavelength range covered by the red CCD is much smaller than that of the blue CCD. The estimated radial velocities and their uncertainties are listed in the last two columns of Table \ref{tab:obs_info}. 

The signal-to-noise ratio of one of the sample stars, HE 0130-1749, is below 20 around the wavelengths of both \ion{K}{1} \qty{766} and \qty{769}{\nano\meter} lines, which is insufficient to obtain useful abundance estimates. For HE 0132-2439, only a few Fe absorption lines were detected to obtain a reliable metallicity estimate. We therefore exclude those two stars in the following analysis.

\begin{deluxetable*}{lcccccccccc}\label{tab:obs_info}
\tablecaption{Summary of the observations}
\tablehead{\colhead{Name} & \colhead{RA} & \colhead{DEC} & \colhead{Date} & \colhead{Exptime} & \colhead{Gmag} & \colhead{SN$_{\rm B}$} & \colhead{SN$_{\rm 766}$} & \colhead{SN$_{\rm 769}$} & \colhead{RV} & \colhead{$\sigma_{\rm RV}$}\\\colhead{-} & \colhead{(deg)} & \colhead{(deg)} & \colhead{-} & \colhead{(s)} & \colhead{(mag)} & \colhead{-} & \colhead{-} & \colhead{-} & \colhead{(km/s)} & \colhead{(km/s)}}
\startdata
CS 30339-0073 & $8.717$ & $-36.924$ & 12-11-2016 & 7200 & $14.5$ & 22 & 25 & 22 & $162.95$ & $0.45$ \\
CS 22942-0002 & $11.649$ & $-24.715$ & 12-11-2016 & 3600 & $13.7$ & 37 & 41 & 33 & $-159.32$ & $0.57$ \\
CD-38 245 & $11.65$ & $-37.658$ & 12-11-2016 & 1200 & $11.7$ & 98 & 118 & 104 & $46.97$ & $1.02$ \\
CD-30 298 & $14.683$ & $-30.098$ & 12-11-2016 & 600 & $10.6$ & 120 & 131 & 115 & $28.75$ & $0.15$ \\
HE 0130-1749 & $23.108$ & $-17.572$ & 18-01-2017 & 7200 & $14.6$ & 16 & 17 & 14 & $-117.52$ & $0.21$ \\
HE 0132-2439 & $23.745$ & $-24.404$ & 12-11-2016 & 7200 & $14.6$ & 20 & 22 & 20 & $-19.03$ & $0.22$ \\
BD+44 493 & $36.709$ & $44.964$ & 12-11-2016 & 600 & $8.9$ & 130 & 143 & 118 & $-149.0$ & $0.38$ \\
CS 22189-0009 & $40.427$ & $-13.469$ & 13-11-2016 & 5400 & $13.8$ & 33 & 37 & 33 & $-20.57$ & $0.37$ \\
CS 22963-0004 & $44.193$ & $-4.855$ & 13-11-2016 & 2700 & $14.8$ & 38 & 38 & 34 & $293.54$ & $0.35$ \\
CS 22172-0002 & $48.586$ & $-10.585$ & 13-11-2016 & 1800 & $12.5$ & 72 & 82 & 74 & $252.12$ & $0.42$ \\
HE 0344-0243 & $56.707$ & $-2.569$ & 18-01-2017 & 7200 & $14.9$ & 19 & 20 & 17 & $-111.23$ & $0.25$ \\
SMSS J085924.06-120104.9 & $134.849$ & $-12.018$ & 12-11-2016 & 8100 & $14.0$ & 32 & 39 & 34 & $203.09$ & $0.25$ \\
HE 0926-0546 & $142.366$ & $-5.996$ & 13-11-2016 & 8100 & $14.0$ & 26 & 30 & 26 & $148.87$ & $1.11$ \\
 HE  1012-1540 & $153.722$ & $-15.934$ & 18-01-2017 & 7200 & $13.9$ & 38 & 37 & 33 & $222.59$ & $0.55$ \\
 BS  16076-0006 & $192.094$ & $20.944$ & 18-01-2017 & 3600 & $13.3$ & 36 & 35 & 30 & $206.3$ & $0.39$ \\
 BS  16929-0005 & $195.872$ & $33.851$ & 18-01-2017 & 4800 & $13.4$ & 42 & 43 & 35 & $-52.17$ & $0.36$ \\
 SDSS J134338.67+484426.6 & $205.911$ & $48.739$ & 18-01-2017 & 2700 & $12.0$ & 111 & 106 & 85 & $-133.07$ & $0.29$ \\
 BS  16550-087 & $212.609$ & $18.022$ & 18-01-2017 & 4800 & $13.5$ & 34 & 36 & 30 & $-148.95$ & $0.29$ \\
CS 22950-0046 & $305.368$ & $-13.275$ & 13-11-2016 & 7200 & $13.9$ & 25 & 32 & 28 & $107.33$ & $0.31$ \\
CS 22949-0048 & $351.53$ & $-5.834$ & 13-11-2016 & 5400 & $13.4$ & 47 & 56 & 46 & $-161.64$ & $0.22$ \\
\enddata

\tablecomments{Basic properties and observational details of the sample stars. The table presents the star names, coordinates in the ICRS, exposure times, G-band magnitudes, signal-to-noise ratios (S/N) for the blue CCD and at the wavelengths of the \ion{K}{1} 766 \si{nm} and 769 \si{nm} lines, as well as the measured radial velocities (RV) and their uncertainties.}\end{deluxetable*}

\subsection{Data reduction}

The raw data were reduced by the software \texttt{hdsql}\footnote{\url{https://github.com/chimari/hds_iraf}}, which utilizes standard \texttt{IRAF} routines \citep{Tody_1993ASPC...52..173T} together with the \texttt{PyRAF} package \citep{PyRAF_2012ascl.soft07011S}.  
Using \texttt{hdsql}, we performed overscan subtraction, linearity correction, cosmic-ray rejection, scattered light subtraction, flat fielding, aperture extraction, wavelength calibration, and heliocentric radial velocity correction.

Special care was taken for the two echelle orders that contained the \ion{K}{1} lines on the red CCD. For those wavelengths, we divide the target spectrum by a standard star spectrum after shifting and rescaling it to minimize the difference between the two spectra in the vicinity of each of the \ion{K}{1} lines. The spectra of standard stars were typically taken within a few hours of the observing time of a given target star. The \texttt{IRAF} routine \texttt{telluric} was used for this purpose.

\section{Abundance analysis}\label{sec:abundance_analysis}

To perform the abundance analysis, we conducted line-by-line spectral synthesis using the \texttt{Turbospectrum} code \citep{Plez_2012ascl.soft05004P} with the MARCS model atmospheres \citep{Gustafsson_2008A&A...486..951G} to fit the observed spectra. 
To automate the processes of interpolating the model atmosphere, generating synthetic spectra, and fitting the synthetic spectra, 
we utilized the \texttt{iSpec} Python interface\citep{Blanco-Cuaresma_2014A&A...569A.111B,Blanco-Cuaresma_2019MNRAS.486.2075B}. 

\subsection{Stellar parameters}

The stellar effective temperature ($T_{\rm eff}$), surface gravity ($\log g$), metallicity ([Fe/H]), and microturbulence velocity ($v_{\rm mic}$) were obtained by the following steps: 

\begin{enumerate}
    \item An initial guess of [Fe/H] values was adopted from the literature.
    \item $T_{\rm eff}$ was estimated based on Gaia DR3 and 2MASS photometry, taking into account the Galactic extinction at estimated distances of the sample stars. 
    \item $\log g$ was determined based on the absolute $G$-band magnitudes calculated based on an extinction-corrected $G$-band magnitude, the $T_{\rm eff}$ estimated by step 1, and a stellar isochrone model.
    \item Microturbulent velocities ($v_{\rm mic}$) were estimated using an empirical formula. Macroturbulent velocity was fixed at \qty{4}{\kilo\meter\per\second} for red giant branch (RGB) stars and \qty{2}{\kilo\meter\per\second} for the main-sequence star. 
    \item The [Fe/H] values were updated based on the Fe abundances from the spectra in this work by adopting the stellar parameters computed in the previous steps.  
\end{enumerate}

In Step 1, we adopted [Fe/H] values from high-resolution spectroscopic analyses of a large sample of very metal-poor stars by \citet{Roderer_2014AJ....147..136R} when available, and supplemented these with values from other literature sources (see Table \ref{tab:stellarpms}).


In Step 2, to accurately estimate the Galactic extinction along the lines of sight to the sample stars, we employed the three-dimensional dust map implemented in the \texttt{dustmaps} Python package \citep{Green2018} to obtain $E(B-V)$ values, incorporating the estimated distances from \citet{Bailer-Jones_2021AJ....161..147B}. For stars located at distances greater than $1$ kpc or at Galactic latitudes $b>60^\circ$, which are likely located beyond the bulk of the dust in the Galactic disk, we adopted extinction values from the two-dimensional dust map of \citet{Schlafly2011ApJ...737..103S}. In all other cases, we used the three-dimensional dust map of \citet{Green2019ApJ...887...93G}.
The estimated extinction values, together with Gaia DR3 and 2MASS photometry, were used to determine $T_{\rm eff}$ based on the color-temperature calibration of \citet{Casagrande_2021MNRAS.507.2684C}, as implemented in the \texttt{colte}\footnote{https://github.com/casaluca/colte} package. The resulting uncertainties in $T_{\rm eff}$ ranged from \qty{62}{\kelvin} to \qty{107}{\kelvin}.

\begin{figure}[htbp]
    \centering
    \includegraphics[width=0.95\linewidth]{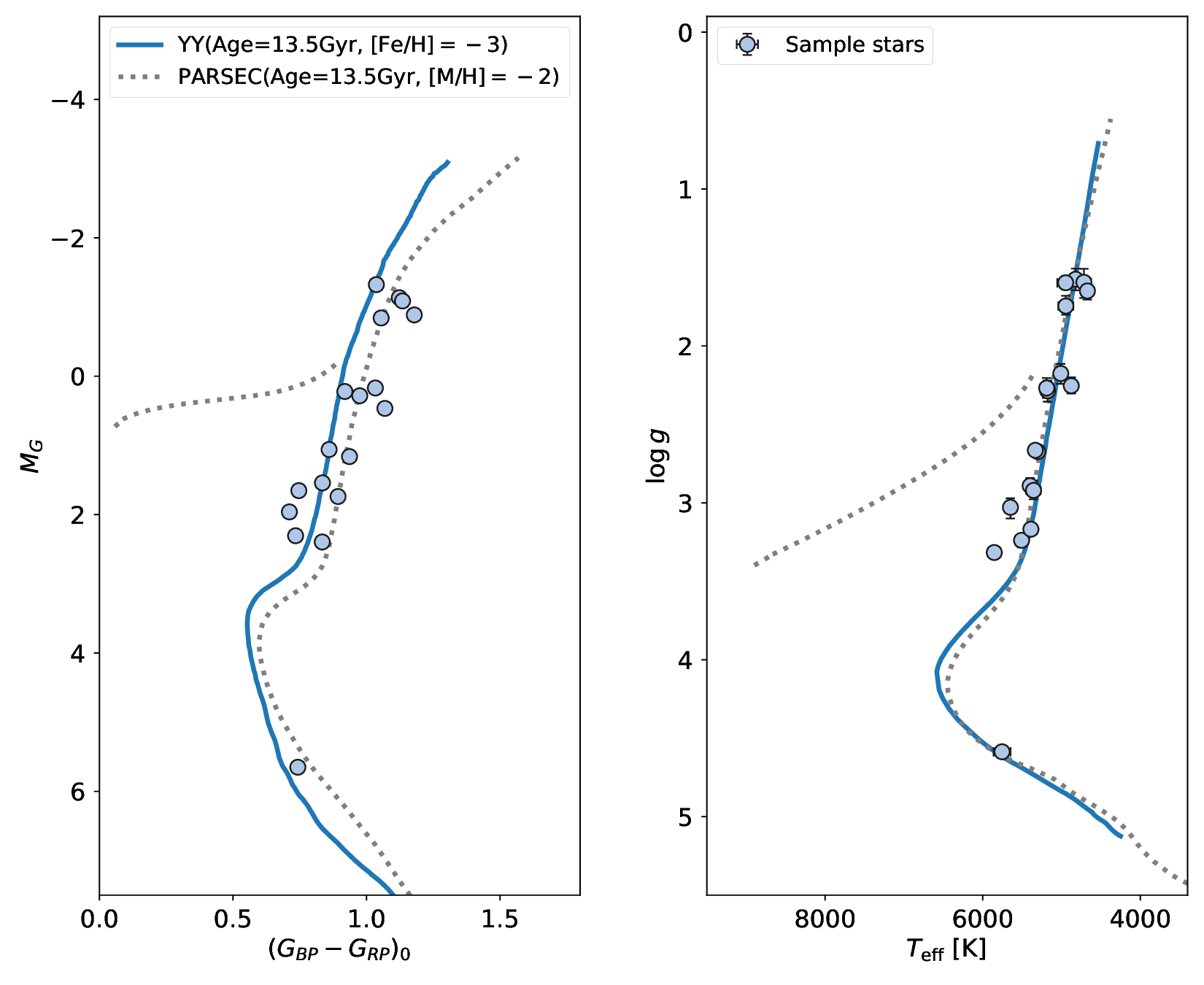}
    \caption{{\it Left}: color-magnitude diagram for the sample stars in the Gaia pass bands (filled circles). The vertical axis shows absolute magnitude in $G$-band, and the horizontal axis shows extinction-corrected $G_{BP}-G_{RP}$ color. {\it Right}: Adopted stellar parameters ($T_{\rm eff}$ and $\log g$) computed in this work. For both panels, 
     the $Y^2$ isochrone model for an age of 13 Gyrs and [Fe/H]$=-3$ used to derive $\log g$ is shown by the solid blue line. For comparison, the PARSEC isochrone for an old ($\log$(Age/yr)=10.13) and metal-poor ([M/H]$=-2$) stellar population is shown as a dotted gray line.}
    \label{fig:stellar_params}
\end{figure}

In Step 3, we first estimated the extinction-corrected absolute $G$-band magnitude ($M_G$) and the $G_{BP}-G_{RP}$ color. The extinction values for the $G$, $G_{BP}$, and $G_{RP}$ bands were derived using the coefficients provided by \citet{Martin2024A&A...692A.115M}, based on the estimated $E(B-V)$, $T_{\rm eff}$, and an initial guess of [Fe/H].  The left panel of Figure~\ref{fig:stellar_params} displays the derived $M_G$ and the extinction-corrected $G_{BP}-G_{RP}$ color.  Bolometric magnitudes ($M_{\rm bol}$) were obtained by interpolating the Yonsei-Yale ($Y^2$) isochrones Version 3 \citep{Demarque2004ApJS..155..667D,Yi2008IAUS..252..413Y} for an age of 13 Gyrs and [Fe/H]$=-3$, at the corresponding $M_G$. The solid blue line represents the $Y^2$ isochrone used for the interpolation. For comparison, the PARSEC isochrone for an old ($\log(\mathrm{Age}/\mathrm{yr}) = 10.13$) and metal-poor ([M/H]$=-2$) stellar population is shown as a dotted gray line \citep{Nguyen_2022A&A...665A.126N}. 
As shown in this figure, the majority of the sample stars are located on the red giant branch, while one star, HE~1012$-$1540, lies on the main sequence. 

Finally, we applied the fundamental relation from \citep{Nissen_1997ESASP.402..225N}, using the estimated values of $M_{\rm bol}$ and $T_{\rm eff}$. 
\begin{equation}\label{eq:logg}
    \log\frac{g}{g_\odot} = \log \frac{M}{M_\odot} + 4\log \frac{T_{\rm eff}}{T_{\rm eff, \odot}} + 0.4(M_{bol} - M_{bol,\odot})
\end{equation}
The adopted solar values are $T_{\rm eff, \odot}=\qty{5777}{\kelvin}$, $\log g_{\odot} =4.4$, 
$M_{bol, \odot}=4.71$. 
Stellar masses in Eq. \ref{eq:logg} are assumed to be 0.71$M_\odot$ for RGB stars and 0.66$M_\odot$ for main sequence stars, which correspond to the median value of the isochrone model in the range $3<M_G<-1$ (RGB) or $5<M_G<6$ (main sequence), respectively. The uncertainties in $\log g$ due to the uncertainty in the distance estimate given by \citet{Bailer-Jones_2021AJ....161..147B} are \qty{0.1}{dex} at most. 

In Step 4, we adopted the empirical formula of \citet{Holtzman_2018AJ....156..125H} to estimate the microturbulent velocity ($v_{\rm mic}$) based on the given values of $\log g$ and [Fe/H]. Minor adjustments were applied, when necessary, to reproduce the observed absorption line profiles as accurately as possible.

In Step 5, the adopted values of [Fe/H] were iteratively updated until convergence within 0.1 dex was achieved. 

The right panel of Figure \ref{fig:stellar_params} shows $T_{\rm eff}$ and $\log g$ as determined in this study. As shown in the figure, these stellar atmospheric parameters are approximately consistent with the adopted $Y^2$ isochrone model.  The $\log g$ values estimated using the $Y^2$ and PARSEC isochrone models agree to within 0.08 dex.

\clearpage

\begin{rotatetable}
\centerwidetable
\begin{deluxetable*}{lccccccccccccccc}
\tablecaption{Stellar parameters}\label{tab:stellarpms}
\tablehead{\colhead{Name} & \colhead{$G$} & \colhead{$G_{\rm BP}$} & \colhead{$G_{\rm RP}$} & \colhead{$M_{G}$} & \colhead{$T_{\rm eff}$} & \colhead{$\sigma_{T_{\rm eff}}$} & \colhead{$\log g$} & \colhead{$\log g_{\rm low}$} & \colhead{$\log g_{\rm high}$} & \colhead{$[{\rm Fe/H}]$} & \colhead{$\sigma_{\rm [Fe/H]}$} & \colhead{$T_{\rm eff}$(ini)} & \colhead{$\log g$(ini)} & \colhead{$[{\rm Fe/H}]$(ini)} & \colhead{Ref}\\\colhead{-} & \colhead{(mag)} & \colhead{(mag)} & \colhead{(mag)} & \colhead{(mag)} & \colhead{(K)} & \colhead{(K)} & \colhead{(dex)} & \colhead{(dex)} & \colhead{(dex)} & \colhead{(dex)} & \colhead{(dex)} & \colhead{(K)} & \colhead{(dex)} & \colhead{(dex)} & \colhead{-}}
\startdata
BS 16076-0006 & $ 13.3 $ & $ 13.7 $ & $ 12.8 $ & $ 2.4 $ & $ 5507 $ & $ 80 $ & $ 3.24 $ & $ 3.26 $ & $ 3.19 $ & $ -3.34 $ & $ 0.03 $ & $ 5199 $ & $ 3.00 $ & $ -3.81 $ & B09 \\
BS 16550-087 & $ 13.5 $ & $ 14.0 $ & $ 12.8 $ & $ -1.1 $ & $ 4821 $ & $ 84 $ & $ 1.58 $ & $ 1.64 $ & $ 1.51 $ & $ -3.34 $ & $ 0.01 $ & $ 4750 $ & $ 1.30 $ & $ -3.39 $ & C13 \\
BS 16929-0005 & $ 13.4 $ & $ 13.8 $ & $ 12.9 $ & $ 1.1 $ & $ 5301 $ & $ 76 $ & $ 2.67 $ & $ 2.71 $ & $ 2.64 $ & $ -3.22 $ & $ 0.01 $ & $ 5245 $ & $ 2.70 $ & $ -3.34 $ & L08 \\
HE 1012-1540 & $ 13.9 $ & $ 14.2 $ & $ 13.3 $ & $ 5.7 $ & $ 5754 $ & $ 107 $ & $ 4.59 $ & $ 4.59 $ & $ 4.58 $ & $ -3.22 $ & $ 0.03 $ & $ 5520 $ & $ 4.70 $ & $ -3.51 $ & C13 \\
SDSSJ134338.67+484426.6 & $ 12.0 $ & $ 12.3 $ & $ 11.6 $ & $ 2.3 $ & $ 5853 $ & $ 80 $ & $ 3.32 $ & $ 3.32 $ & $ 3.31 $ & $ -3.10 $ & $ 0.01 $ & $ 6030 $ & $ 3.34 $ & $ -3.15 $ & L21 \\
BD+44 493 & $ 8.9 $ & $ 9.2 $ & $ 8.3 $ & $ 2.0 $ & $ 5388 $ & $ 82 $ & $ 3.17 $ & $ 3.17 $ & $ 3.16 $ & $ -3.65 $ & $ 0.01 $ & $ 5040 $ & $ 2.10 $ & $ -4.28 $ & R14 \\
CD-30 298 & $ 10.6 $ & $ 11.0 $ & $ 10.0 $ & $ 1.2 $ & $ 5332 $ & $ 76 $ & $ 2.67 $ & $ 2.68 $ & $ 2.65 $ & $ -3.09 $ & $ 0.01 $ & $ 4810 $ & $ 1.50 $ & $ -3.77 $ & R14 \\
CD-38 245 & $ 11.7 $ & $ 12.2 $ & $ 11.0 $ & $ -1.3 $ & $ 4947 $ & $ 107 $ & $ 1.60 $ & $ 1.64 $ & $ 1.55 $ & $ -3.81 $ & $ 0.01 $ & $ 4520 $ & $ 0.65 $ & $ -4.59 $ & R14 \\
CS 22189-0009 & $ 13.8 $ & $ 14.2 $ & $ 13.2 $ & $ 0.2 $ & $ 5009 $ & $ 76 $ & $ 2.18 $ & $ 2.24 $ & $ 2.11 $ & $ -3.26 $ & $ 0.01 $ & $ 4540 $ & $ 0.60 $ & $ -3.92 $ & R14 \\
CS 22172-0002 & $ 12.5 $ & $ 12.9 $ & $ 11.8 $ & $ -0.8 $ & $ 4945 $ & $ 94 $ & $ 1.75 $ & $ 1.81 $ & $ 1.69 $ & $ -3.53 $ & $ 0.01 $ & $ 4800 $ & $ 1.30 $ & $ -3.86 $ & H13 \\
CS 22942-0002 & $ 13.7 $ & $ 14.0 $ & $ 13.1 $ & $ 1.5 $ & $ 5398 $ & $ 71 $ & $ 2.89 $ & $ 2.94 $ & $ 2.84 $ & $ -3.03 $ & $ 0.01 $ & $ 5010 $ & $ 2.00 $ & $ -3.61 $ & R14 \\
CS 22949-0048 & $ 13.4 $ & $ 13.9 $ & $ 12.7 $ & $ 0.5 $ & $ 4876 $ & $ 64 $ & $ 2.25 $ & $ 2.31 $ & $ 2.20 $ & $ -3.15 $ & $ 0.01 $ & $ 4620 $ & $ 0.95 $ & $ -3.55 $ & R14 \\
CS 22950-0046 & $ 13.9 $ & $ 14.4 $ & $ 13.2 $ & $ -1.1 $ & $ 4716 $ & $ 73 $ & $ 1.59 $ & $ 1.68 $ & $ 1.49 $ & $ -3.45 $ & $ 0.01 $ & $ 4380 $ & $ 0.50 $ & $ -4.12 $ & R14 \\
CS 22963-0004 & $ 14.8 $ & $ 15.1 $ & $ 14.3 $ & $ 1.7 $ & $ 5648 $ & $ 73 $ & $ 3.03 $ & $ 3.09 $ & $ 2.96 $ & $ -3.03 $ & $ 0.02 $ & $ 5060 $ & $ 2.15 $ & $ -4.09 $ & R14 \\
CS 30339-0073 & $ 14.5 $ & $ 14.9 $ & $ 14.0 $ & $ 1.7 $ & $ 5355 $ & $ 90 $ & $ 2.92 $ & $ 2.98 $ & $ 2.86 $ & $ -3.16 $ & $ 0.02 $ & $ 4830 $ & $ 1.55 $ & $ -3.93 $ & R14 \\
HE 0344-0243 & $ 14.9 $ & $ 15.4 $ & $ 14.3 $ & $ 0.2 $ & $ 5176 $ & $ 77 $ & $ 2.29 $ & $ 2.35 $ & $ 2.22 $ & $ -3.00 $ & $ 0.03 $ & $ 5140 $ & $ 2.30 $ & $ -3.35 $ & C13 \\
HE 0926-0546 & $ 14.0 $ & $ 14.4 $ & $ 13.4 $ & $ 0.3 $ & $ 5188 $ & $ 81 $ & $ 2.27 $ & $ 2.33 $ & $ 2.21 $ & $ -3.55 $ & $ 0.02 $ & $ 5159 $ & $ 2.50 $ & $ -3.73 $ & C13 \\
SMSS J085924.06-120104.9 & $ 14.0 $ & $ 14.6 $ & $ 13.3 $ & $ -0.9 $ & $ 4671 $ & $ 81 $ & $ 1.65 $ & $ 1.73 $ & $ 1.59 $ & $ -3.33 $ & $ 0.01 $ & $ 4459 $ & $ 1.20 $ & $ -3.48 $ & J15 \\
\enddata
\tablecomments{Stellar parameters derived in this study. $G$, $G_{\rm BP}$, $G_{\rm RP}$ represent the appearent magnitudes in the Gaia passbands. $M_{G}$ is the estimated absolute magnitude based on Gaia DR3 parallaxes. The uncertainty in $T_{\rm eff}$ reflects the calibration uncertainty from \citet{Casagrande_2021MNRAS.507.2684C}. The values of $\log g_{\rm low}$ and $\log g_{\rm high}$ indicate the surface gravity estimates when parallaxes are varied within their reported uncertainties. The uncertainty in [Fe/H] includes only statistical errors.}
\end{deluxetable*}

\end{rotatetable}

\clearpage

\subsection{Atomic data}

The adopted atomic data for \ion{K}{1} and other metal absorption lines is shown in Table \ref{tab:linelist}.

For elements other than K, we compiled a line list suitable for extremely metal-poor stars and cover the spectral range of our observational setup from \citet{Ishigaki_2012ApJ...753...64I,Ishigaki_2013ApJ...771...67I}. We supplement the lines from \citet{Yong_2013ApJ...762...26Y} to optimize the line selection for extremely metal-poor stars.  We estimated the abundance of 
Na, Mg, Ca, Ti, Cr, Fe, Ni, while the lines of Si, Sc, Mn, Co, Y, and Ba were too weak in the wavelength range covered in this work. We report the results of Na, Mg, Ca, Ti, Cr, Fe, and Ni in the following sections.
The atomic data were adopted from version 6 of the {\it Gaia}-ESO line list, compiled by \citet{Heiter2021A&A...645A.106H}. Those data are based on various literature sources and the \href{https://vald.astro.uu.se}{\texttt{VALD}} database \citep{Piskunov1995A&AS..112..525P,Kupka1999A&AS..138..119K,Ryabchikova2015PhyS...90e4005R}.

\subsection{Metallicity}

\begin{figure}
    \centering
    \includegraphics[width=0.95\linewidth]{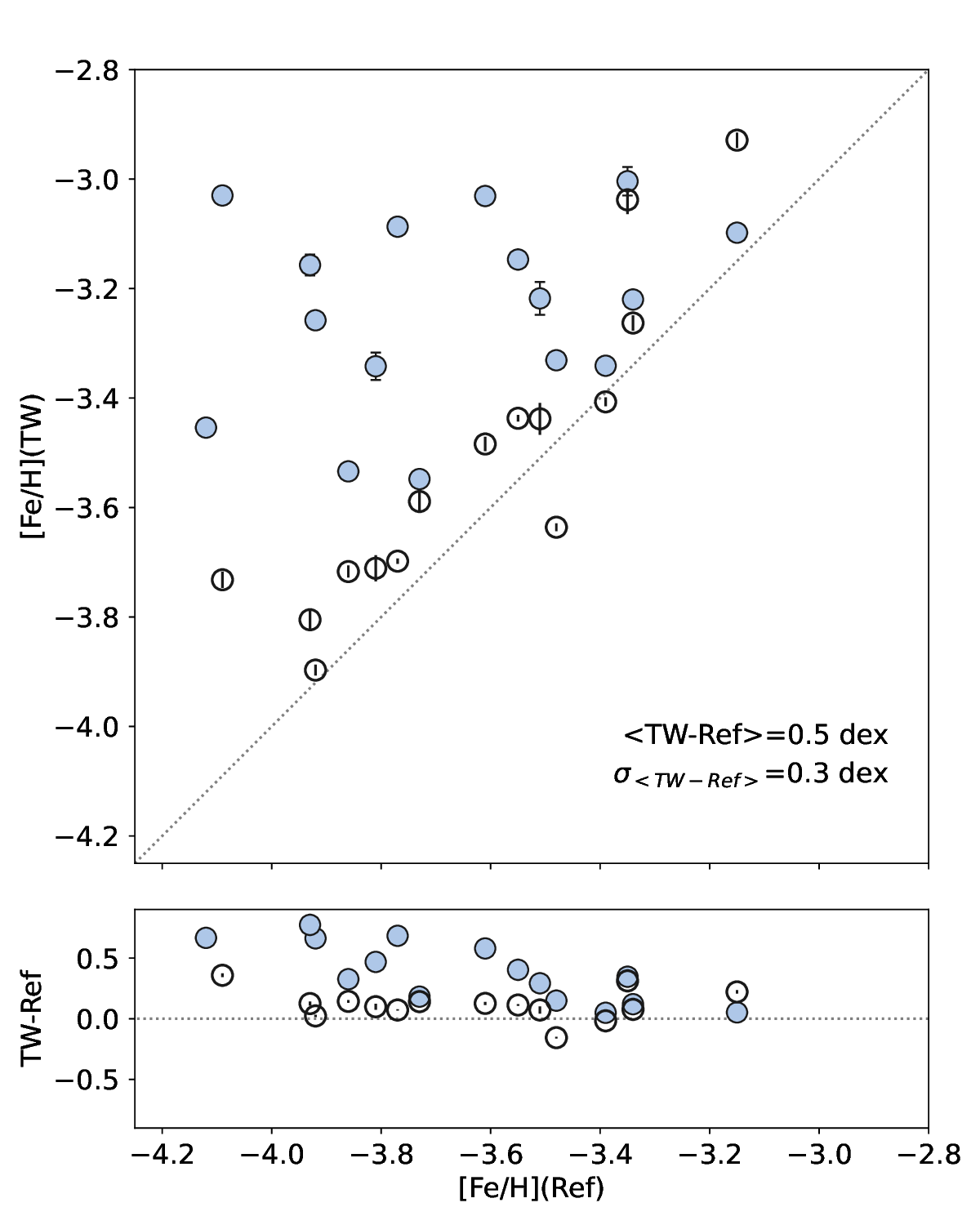}
    \caption{The comparison of [Fe/H] values derived in this work (the vertical axis) plotted against [Fe/H] values from literature (the horizontal axis). Filled blue circles correspond to the values obtained by adopting $T_{\rm eff}$ and $\log g$ in this work. Open gray circles correspond to the values obtained by adopting the stellar parameters from the literature listed in Table \ref{tab:stellarpms}.}
    \label{fig:feh_comparison}
\end{figure}

Figure \ref{fig:feh_comparison} compares the values of [Fe/H] obtained in this work and those from the literature from which initial guesses of [Fe/H] values were adopted (see Table \ref{tab:stellarpms}).  The [Fe/H] values estimated in this work are systematically higher than those derived in previous studies, sometimes by more than \qty{0.5}{dex}. In particular, for 9 stars analyzed in common with \citet{Roderer_2014AJ....147..136R}, the differences reach up to \qty{1.0}{dex}. 
The differences in [Fe/H] estimates are mainly attributed to differences in the temperature scales adopted in this work and in the literature, where \citet{Roderer_2014AJ....147..136R} used \ion{Fe}{1} lines to derive $T_{\rm eff}$ by requiring the Fe abundances to not show a trend with the excitation potentials. The $\log g$ value was then derived using the $Y^2$ isochrone model \citep{Demarque2004ApJS..155..667D}. For CD-38 245 as an example, \citet{Roderer_2014AJ....147..136R} obtained $T_{\rm eff}=$\qty{4520}{\kelvin}, $\log g=0.65$, and [Fe/H]$=-4.59$. In contrast, this study obtained $T_{\rm eff}=$\qty{4947}{\kelvin}, $\log g=1.60$, and [Fe/H]$=-3.82$. In the former case, $\log g$ is incompatible with the distance estimate of $d=$ 3710 pc from \citet{Bailer-Jones_2021AJ....161..147B}. 
The gray open circles in Figure \ref{fig:feh_comparison} show that the [Fe/H]
 values estimated in this work when the same values of $T_{\rm eff}$ and $\log g$ were adopted in the abundance analysis. The agreement is significantly improved in this case.  

To ensure homogeneity in the method by which the values of $T_{\rm eff}$, $\log g$, and [Fe/H] are derived, we adopted these parameters from this work in the following analysis. 
We address the consistency between the abundance ratios derived in this work and those from the literature in Section \ref{sec:other_elem}.

\begin{deluxetable}{lccc}\label{tab:linelist}
\tablecaption{Atomic linelist}
\tablehead{\colhead{Element} & \colhead{Wavelength} & \colhead{Excp} & \colhead{$\log gf$}\\ \colhead{-} & \colhead{nm} & \colhead{eV} & \colhead{-}}
\startdata
K 1 & $ 766.490 $ & $ 0.000 $ & $ 0.149 $ \\
K 1 & $ 769.896 $ & $ 0.000 $ & $ -0.154 $ \\
Na 1 & $ 568.263 $ & $ 2.102 $ & $ -0.706 $ \\
Na 1 & $ 588.995 $ & $ 0.000 $ & $ 0.108 $ \\
Na 1 & $ 589.592 $ & $ 0.000 $ & $ -0.144 $ \\
Na 1 & $ 615.423 $ & $ 2.102 $ & $ -1.547 $ \\
Na 1 & $ 616.075 $ & $ 2.104 $ & $ -1.246 $ \\
Mg 1 & $ 517.268 $ & $ 2.712 $ & $ -0.450 $ \\
Mg 1 & $ 518.360 $ & $ 2.717 $ & $ -0.239 $ \\
\enddata

\tablecomments{Table 1 is published in its entirety in the machine-readable format.
      A portion is shown here for guidance regarding its form and content.}
\end{deluxetable}

\subsection{Potassium}

\begin{figure*}
    \centering
    \plottwo{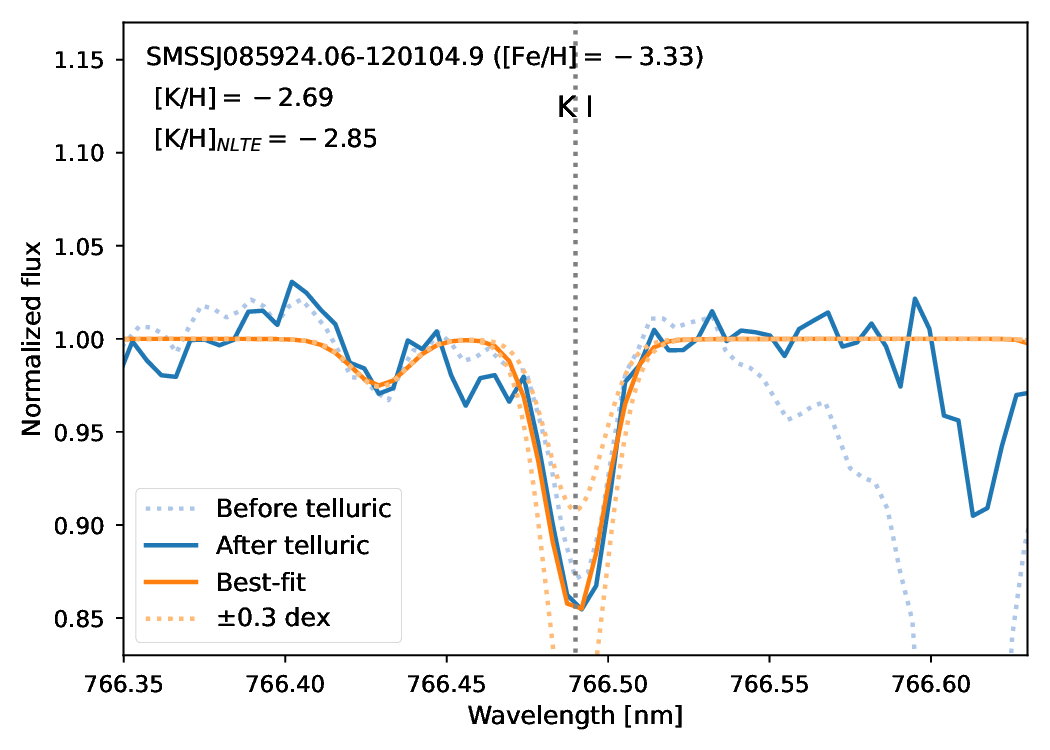}{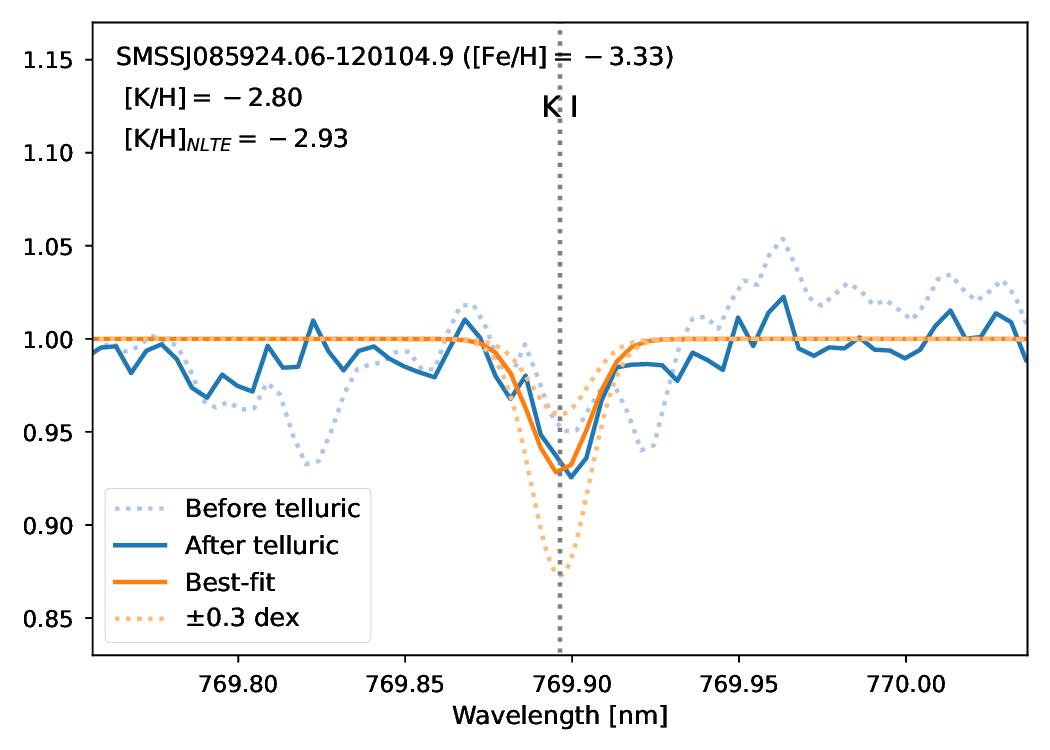}
    \plottwo{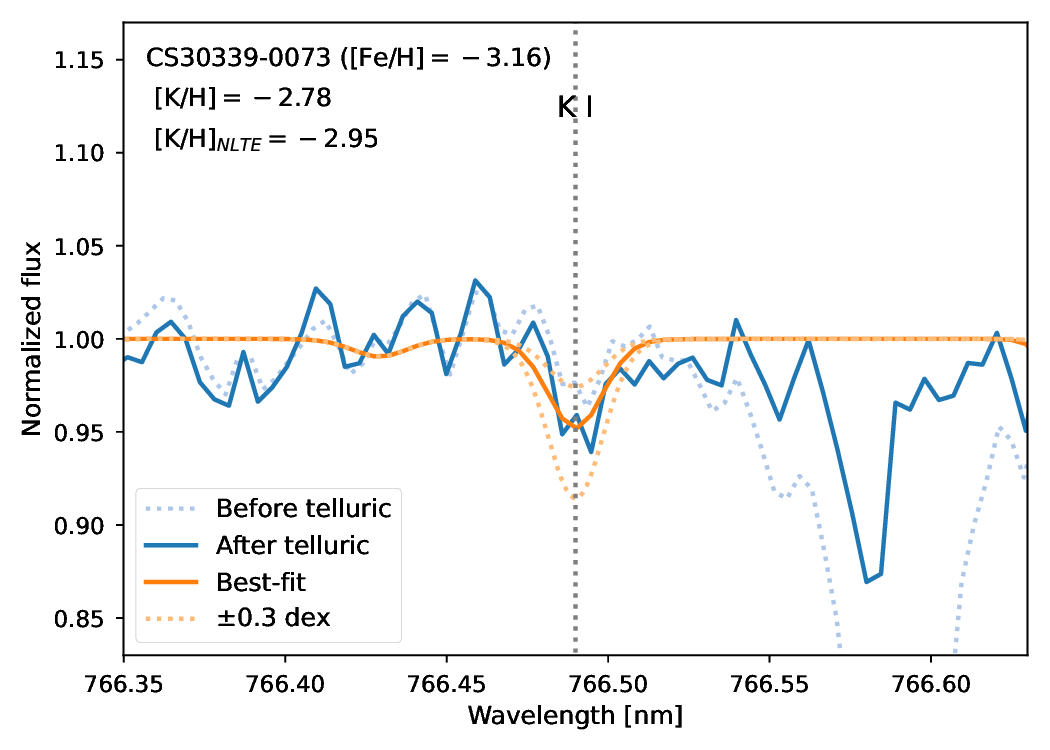}{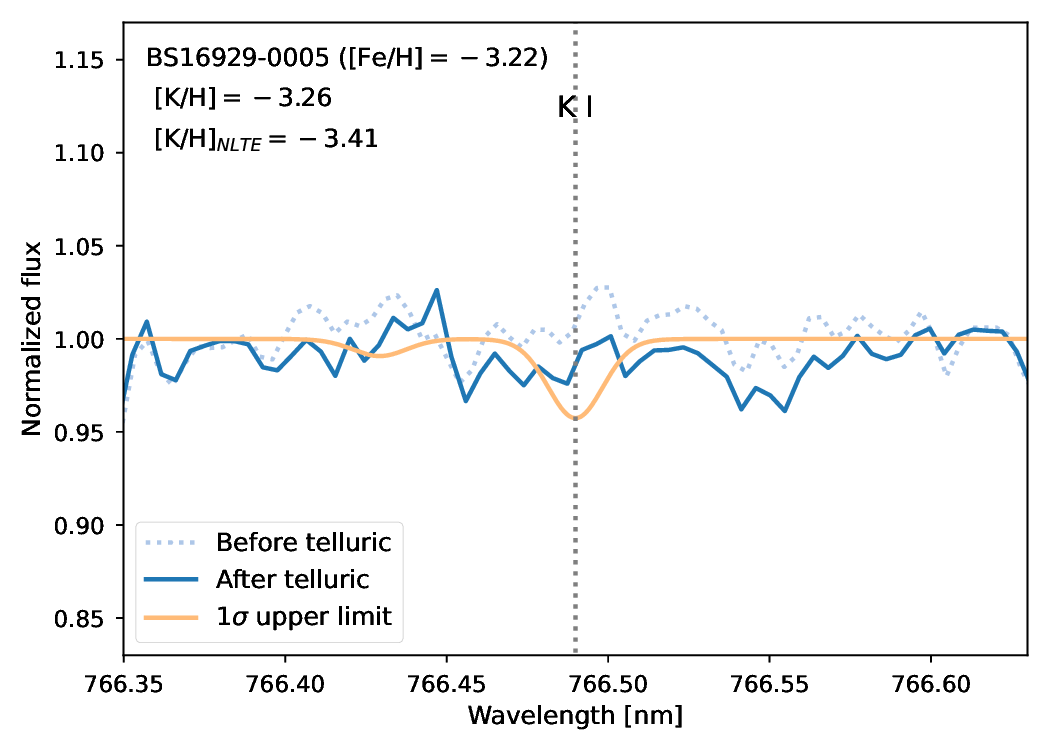}
    \caption{Observed and synthetic spectra for the \ion{K}{1} lines. The light blue and dark blue solid lines represent the observed spectra before and after telluric correction, respectively. The solid orange lines indicate the best-fit synthetic spectra or represent an upper limit. The dotted orange lines show synthetic spectra with K abundances offset by $\pm 1\sigma$.  The top two panels display the two \ion{K}{1} lines for SMSS J085924.06-120104.9. The bottom-left panel shows the spectrum of the lower-metallicity star CS 30339-0073. The bottom-right panel shows the case of BS 16929-0005, for which only an upper limit on the K abundance could be determined.}
    \label{fig:Klines_detected}
\end{figure*}

Figure \ref{fig:Klines_detected} presents examples of observed spectra before and after correcting for telluric absorption using the standard star spectra. The top two panels show the \ion{K}{1} lines at \qty{766.49} and \qty{769.90}{\nano\meter} for one of the sample stars. The bottom-left panel shows the \ion{K}{1} line at \qty{766.49}{\nano\meter} for a star with lower metallicity. As demonstrated in these examples, the dominant contribution from telluric lines was successfully removed when the telluric features were offset from the \ion{K}{1} lines.
Measurements were excluded when a \ion{K}{1} line significantly overlapped with telluric absorption. The bottom-right panel illustrates a case in which only an upper limit on the potassium abundance could be derived.

\subsubsection{Synthetic spectral fitting \label{sec:synfit}}

We fitted synthetic spectra to the observed spectra after normalizing the local continuum. The best-fit spectrum for a representative case is shown in Figure \ref{fig:Klines_detected}. We detected the \ion{K}{1} \qty{766.49}{\nano\meter} line in five of the sample stars, and the \qty{769.90}{\nano\meter} line in five of the sample stars. 
If the detection of the \ion{K}{1} line is deemed uncertain based on visual inspection, we report only upper limits. 
Uncertainty on K abundances is estimated by performing 100 iterations of K abundance measurements using synthetic spectra with added noise, simulating the statistical fluctuations of the observed spectra at a given signal-to-noise ratio. Similarly, the one-sigma upper limits 
on K abundances were estimated by performing the measurements using line-free synthetic spectra. The sum of the mean K abundance derived from line-free spectra and the corresponding one-sigma scatter was reported as the one-sigma upper limit. An example synthetic spectrum corresponding to the one-sigma upper limit is shown in the bottom-right panel of Figure \ref{fig:Klines_detected}.

\subsubsection{NLTE correction}

Both \ion{K}{1} lines used in this study are known to be sensitive to deviations from LTE \citep{Takeda_2002PASJ...54..275T,Andrievsky_2010A&A...509A..88A,Zhao_2016ApJ...833..225Z,Reggiani_2019A&A...627A.177R}. 
We use the grid of NLTE corrections for the \ion{K}{1} \qty{766.49} / \qty{769.90}{\nano\meter} lines calculated by \citet{Reggiani_2019A&A...627A.177R}. The grid is interpolated to obtain corrections appropriate for each sample star, based on its estimated values of $T_{\rm eff}$, $\log g$, [Fe/H], $v_{mic}$, and [K/Fe].

\subsection{Other elements}\label{sec:other_elem}

The same spectral fitting technique described in Section \ref{sec:synfit} was used to determine the abundances of Na, Mg, Ca, Cr, Ti, and Ni. Only a small number of stars exhibited measurable absorption lines for additional elements (e.g., Al, Mn, Zn, etc.), primarily due to the redder wavelength coverage of the spectra analyzed in this study. To better constrain the nucleosynthetic yield models 
that account for the overall abundance patterns of the sample stars
(see Section \ref{sec:origin_K}), we supplemented the abundance estimates of Sc, V, and Mn using values from \citet{Roderer_2014AJ....147..136R}, \citet{Jacobson_2015}, and \citet{Lai_2008ApJ...681.1524L}.

\begin{figure*}[htbp]
\gridline{
\fig{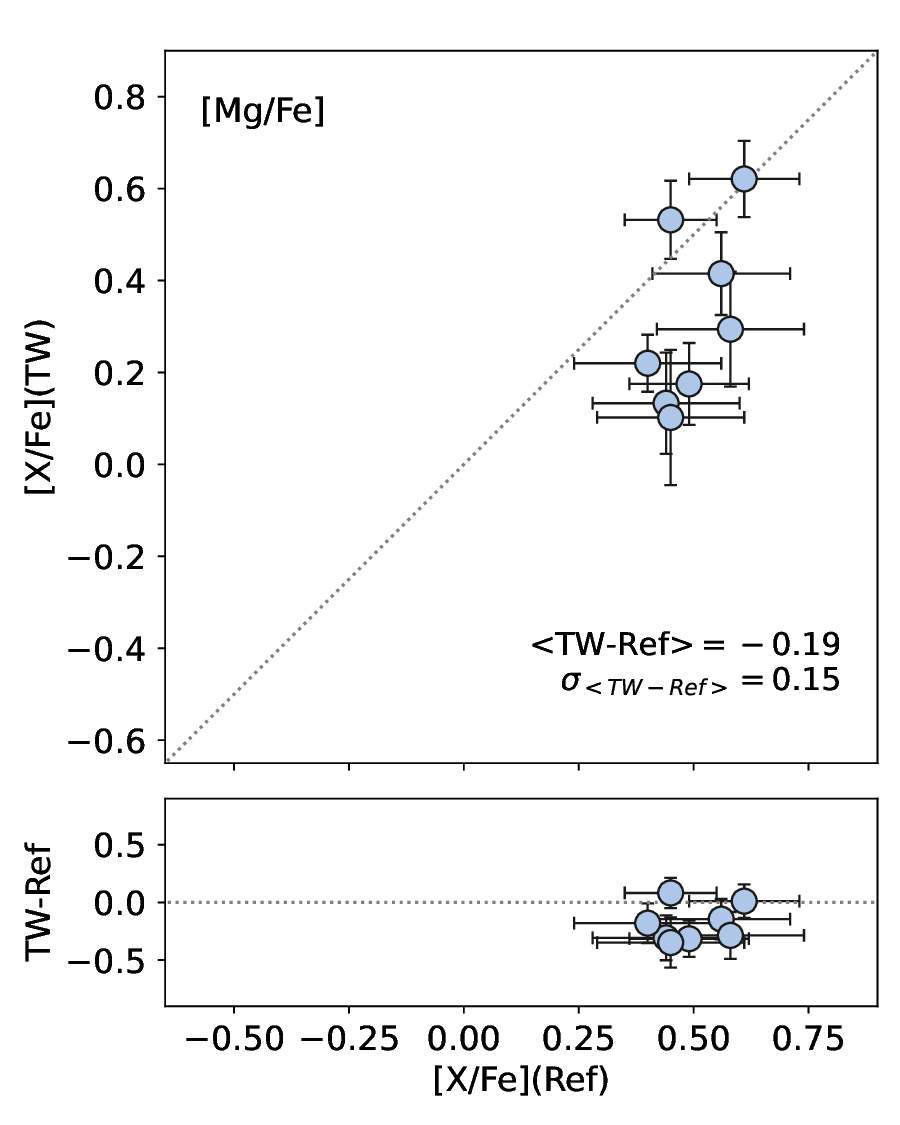}{0.3\textwidth}{(a)}
\fig{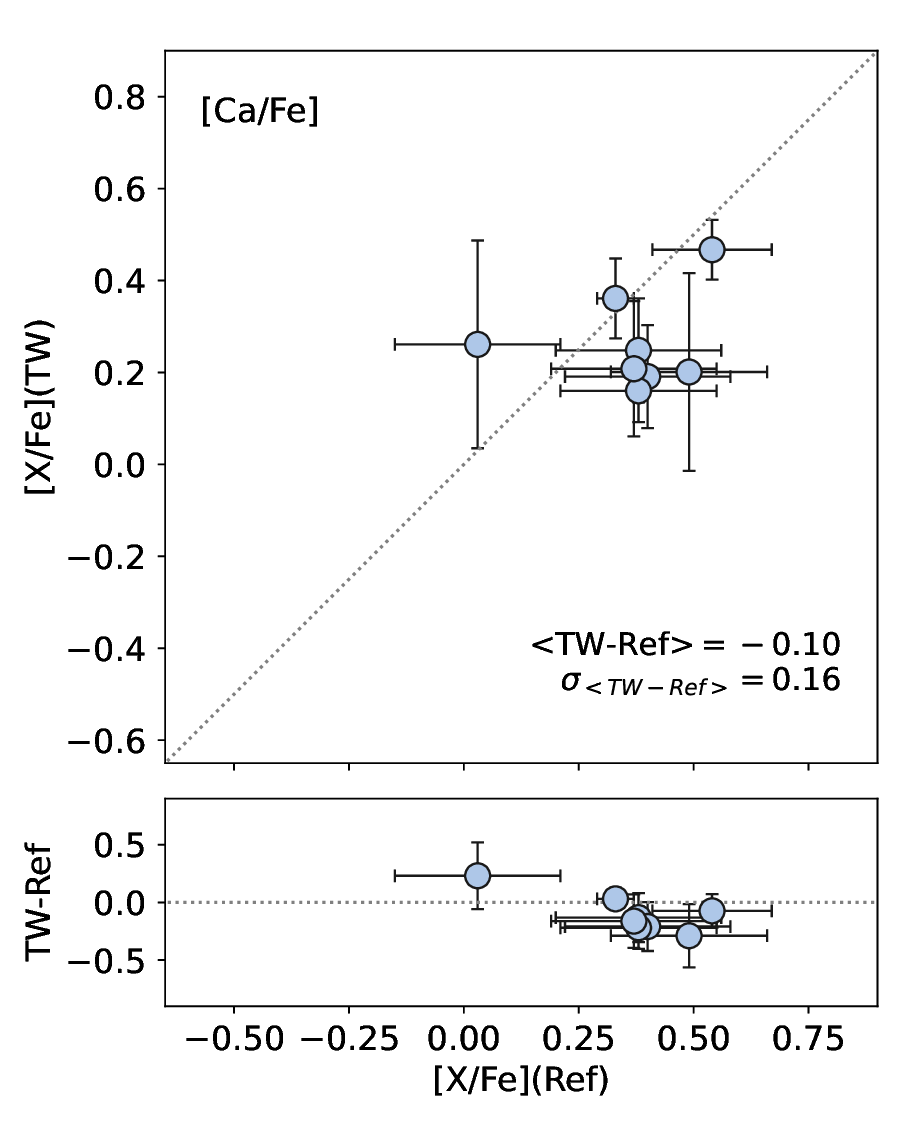}{0.3\textwidth}{(b)}
\fig{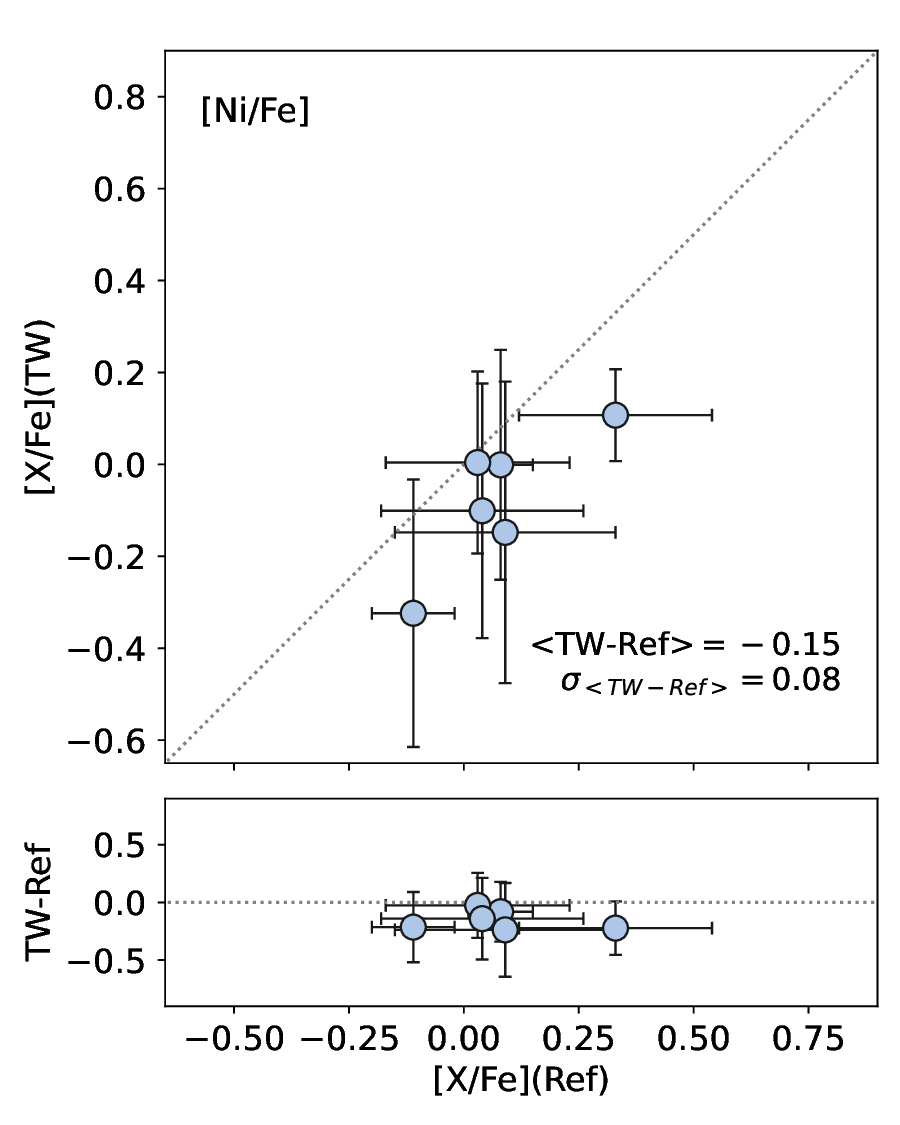}{0.3\textwidth}{(c)}
}
\caption{The comparison of [Mg/Fe] (a), [Ca/Fe](b), and [Ni/Fe] (c) ratios derived in this work (the vertical axis) plotted against corresponding values from literature listed in Table \ref{tab:stellarpms} (the horizontal axis).} 
\label{fig:elem_comparison}
\end{figure*}

Figure \ref{fig:elem_comparison} compares the [Mg/Fe], [Ca/Fe], and [Ni/Fe] ratios obtained in this work and those from the literature (Table \ref{tab:stellarpms}). Despite the large differences in the adopted [Fe/H] values (see Figure \ref{fig:feh_comparison}), the mean offsets of the abundance ratios are up to \qty{0.19}{dex}. 
The mean offset of \qty{-0.15}{dex} in the [Ni/Fe] ratios is significantly larger than the standard deviation of the differences. The discrepancy between the Ni abundance estimates in this work and those from \citet{Roderer_2014AJ....147..136R}, which are primarily adopted as literature values, is likely due to the different 
sets of \ion{Ni}{1} lines used. While \citet{Roderer_2014AJ....147..136R} mainly used lines at shorter wavelengths ($<$\qty{500}{\nano\meter}), this study utilizes lines at longer wavelengths. In \citet{Roderer_2014AJ....147..136R}, one \ion{Ni}{1} line at \qty{547.7}{\nano\meter}, which 
is commonly analyzed in both studies, yields 
lower Ni abundances compared to those derived from the shorter-wavelength 
lines. This systematic difference does not directly affect the main conclusions of this work.

\section{Result}\label{sec:result}

\subsection{[K/Fe], [K/Ca] ratios}\label{sec:kfe_kca}

Among the 18 stars analyzed, K abundances were estimated for seven stars based on at least one of the \ion{K}{1} lines. For the other eleven objects, only a $1\sigma$ upper limit could be determined. 
The K abundances derived have been summarized in Table \ref{tab:K_abundance}.

\begin{figure*}[htbp]
    \centering
    \includegraphics[width=0.95\linewidth]{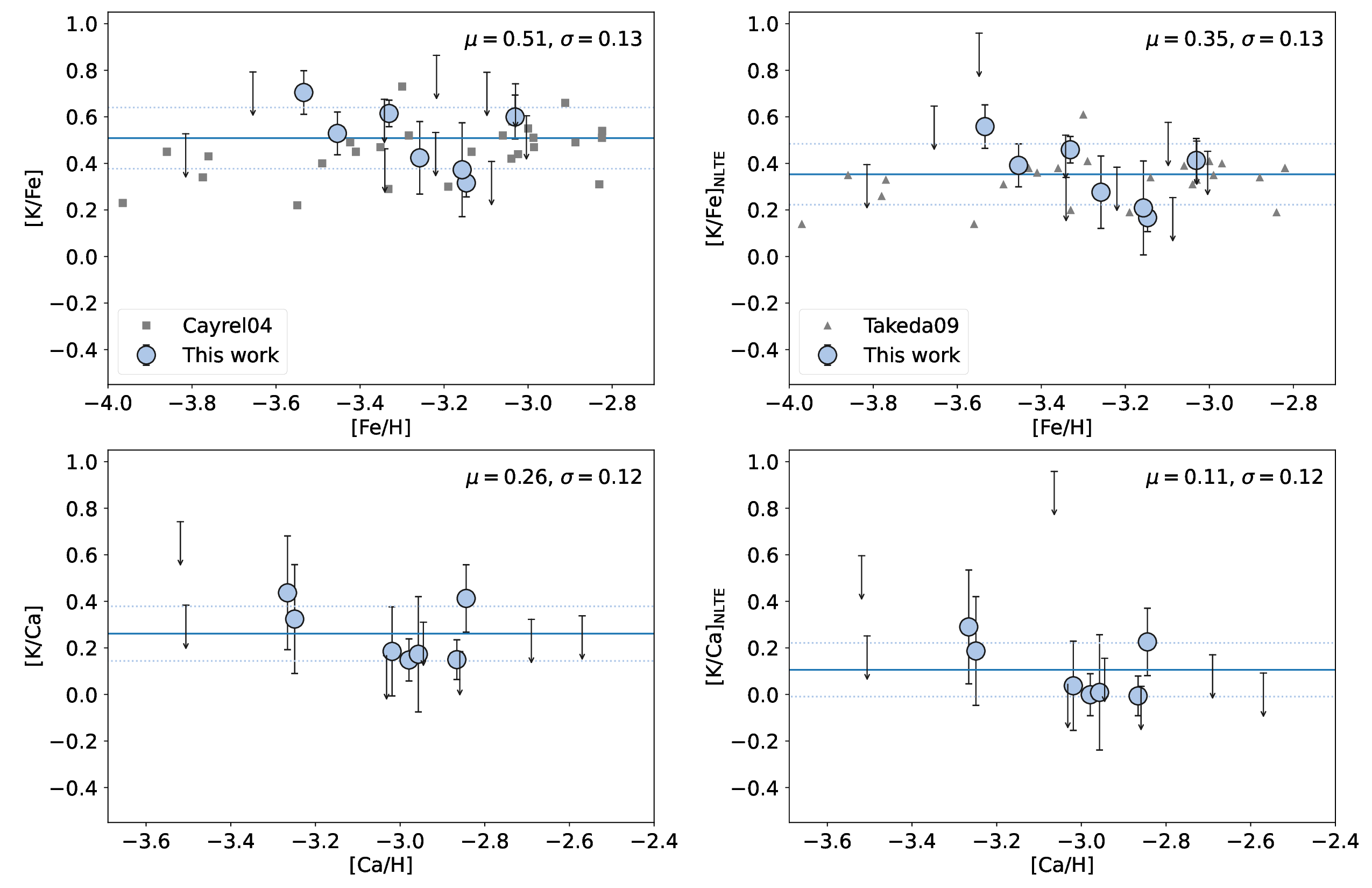}
    \caption{[K/Fe] and [K/Ca] ratios from the LTE (left panels) and NLTE (right panels) analyses obtained in this work. The stars with one or both \ion{K}{1} lines detected are shown as circles with error bars. The stars with only an upper limit are shown by arrows, where the location of the arrowheads indicates the 1 $\sigma$ level of the noise in the spectrum.  For comparison, the abundances derived by \citet{Cayrel_2004A&A...416.1117C} and \citet{Takeda_2009PASJ...61..563T} are shown by gray squares and triangles, respectively. Horizontal lines show the mean and standard deviation of the abundance ratios for the stars with detected \ion{K}{1} lines.} 
    \label{fig:kfe_kca}
\end{figure*}

The resulting [K/Fe] values from the LTE analysis are shown in the top-left panel of Figure \ref{fig:kfe_kca}.  For the seven stars with measured [K/Fe] abundances, the values range from \qty{0.3} to \qty{0.7}{dex}, with a mean of \qty{0.51}{dex} and a standard deviation of \qty{0.13}{dex}. This dispersion is smaller than the typical statistical uncertainty in the measured [K/Fe] abundances (approximately \qty{0.2}{dex}). The upper limits obtained for the remaining stars are consistent with the 2$\sigma$ range of the measured abundances. For comparison, LTE abundances from \citet{Cayrel_2004A&A...416.1117C}
are also plotted in the same panel. The [K/Fe] ratios derived in this study are in good agreement with those reported by \citet{Cayrel_2004A&A...416.1117C}.

The top-right panel of Figure \ref{fig:kfe_kca} shows the [K/Fe] ratios after applying the NLTE corrections based on the calculations by \citet{Reggiani_2019A&A...627A.177R}. The mean [K/Fe] value is \qty{0.35}{dex}, which is lower than the LTE result by a difference of \qty{0.16}{dex}.  The standard deviation remains \qty{0.13}{dex}, again less than the typical statistical uncertainty. It should be noted that systematic uncertainties associated with the NLTE corrections are not included in this analysis. For comparison, NLTE K abundances from \citet{Takeda_2009PASJ...61..563T}, based on a re-analysis of the \citet{Cayrel_2004A&A...416.1117C} sample, are also shown. 
The [K/Fe] ratios derived in this study are consistent with the super-solar values and small scatter observed in the sample of \citet{Takeda_2009PASJ...61..563T} and \citet{Reggiani_2019A&A...627A.177R}. 
The low upper limit for CD-38 245 at [Fe/H]$=-3.8$ is consistent with the slightly decreasing trend of [K/Fe]$_{\rm NLTE}$ seen in the sample of \citet{Takeda_2009PASJ...61..563T}.

The two bottom panels of Figure \ref{fig:kfe_kca} present the [K/Ca] and [K/Ca]$_{\rm NLTE}$ ratios derived from LTE and NLTE analyses, plotted as a function of [Ca/H]. Using Ca as the reference element instead of Fe provides a clearer insight into the process of K nucleosynthesis, as both elements originate from similar layers within the progenitor star before and after the core-collapse supernova explosion. 
For the stars with detected K lines, the mean [K/Ca]$_{\rm NLTE}$ ratio is \qty{0.11}{dex}, with a scatter of \qty{0.12}{dex}.

\begin{deluxetable*}{lcccccccccccc}\label{tab:K_abundance}
\tablecaption{K abundances}
\tablehead{\colhead{Name} & \colhead{[Fe/H]} & \colhead{[Ca/H]} & \colhead{} & \colhead{[K/H](766)} & \colhead{$\Delta_{\rm NLTE}$(766)} & \colhead{} & \colhead{[K/H](769)} & \colhead{$\Delta_{\rm NLTE}$(769)} & \colhead{} & \colhead{[K/Fe]} & \colhead{e[K/Fe]} & \colhead{[K/Fe]$_{
m NLTE}$}\\\colhead{-} & \colhead{(dex)} & \colhead{(dex)} & \colhead{-} & \colhead{(dex)} & \colhead{(dex)} & \colhead{-} & \colhead{(dex)} & \colhead{(dex)} & \colhead{-} & \colhead{(dex)} & \colhead{(dex)} & \colhead{(dex)}}
\startdata
CS 22189-0009 & $ -3.26 $ & $ -3.02 $ &  & - & - &  & $ -2.83 $ & $ -0.15 $ &  & $ 0.42 $ & $ 0.16 $ & $ 0.28 $ \\
CS 22942-0002 & $ -3.03 $ & $ -2.84 $ &  & $ -2.43 $ & $ -0.19 $ & $<$ & $ -2.57 $ & $ -0.17 $ &  & $ 0.60 $ & $ 0.09 $ & $ 0.41 $ \\
CS 22172-0002 & $ -3.53 $ & $ -3.27 $ &  & - & - &  & $ -2.83 $ & $ -0.15 $ &  & $ 0.70 $ & $ 0.09 $ & $ 0.56 $ \\
SMSS J085924.06-120104.9 & $ -3.33 $ & $ -2.87 $ &  & $ -2.69 $ & $ -0.16 $ &  & $ -2.80 $ & $ -0.13 $ &  & $ 0.61 $ & $ 0.06 $ & $ 0.46 $ \\
CS 30339-0073 & $ -3.16 $ & $ -2.96 $ &  & $ -2.78 $ & $ -0.16 $ &  & - & - &  & $ 0.37 $ & $ 0.20 $ & $ 0.21 $ \\
CS 22950-0046 & $ -3.45 $ & $ -3.25 $ &  & $ -2.90 $ & $ -0.14 $ &  & $ -2.99 $ & $ -0.13 $ &  & $ 0.53 $ & $ 0.09 $ & $ 0.39 $ \\
CS 22949-0048 & $ -3.15 $ & $ -2.98 $ &  & $ -2.83 $ & $ -0.15 $ &  & $ -2.83 $ & $ -0.14 $ &  & $ 0.32 $ & $ 0.06 $ & $ 0.17 $ \\
BS 16929-0005 & $ -3.22 $ & $ -2.86 $ & $<$ & $ -3.26 $ & $ -0.15 $ &  & - & - & $<$ & $ 0.25 $ & $ 0.29 $ & $ 0.10 $ \\
BS 16550-087 & $ -3.34 $ & $ -3.03 $ &  & - & - & $<$ & $ -3.42 $ & $ -0.12 $ & $<$ & $ 0.19 $ & $ 0.27 $ & $ 0.07 $ \\
BS 16076-0006 & $ -3.34 $ & $ -3.44 $ &  & - & - & $<$ & $ -3.18 $ & $ -0.15 $ & $<$ & $ 0.42 $ & $ 0.25 $ & $ 0.27 $ \\
HE 1012-1540 & $ -3.22 $ & $ -2.98 $ & $<$ & $ -3.00 $ & - & $<$ & $ -2.40 $ & - & $<$ & $ 0.54 $ & $ 0.33 $ & - \\
CD-38 245 & $ -3.81 $ & $ -3.51 $ & $<$ & $ -3.50 $ & $ -0.13 $ & $<$ & $ -3.22 $ & $ -0.13 $ & $<$ & $ 0.42 $ & $ 0.11 $ & $ 0.29 $ \\
SDSSJ134338.67+484426.6 & $ -3.10 $ & $ -3.09 $ &  & - & - & $<$ & $ -2.54 $ & $ -0.22 $ & $<$ & $ 0.68 $ & $ 0.12 $ & $ 0.46 $ \\
BD+44 493 & $ -3.65 $ & $ -3.52 $ & $<$ & $ -3.03 $ & $ -0.15 $ & $<$ & $ -3.22 $ & $ -0.15 $ & $<$ & $ 0.61 $ & $ 0.18 $ & $ 0.47 $ \\
CD-30 298 & $ -3.09 $ & $ -2.94 $ &  & - & - & $<$ & $ -2.89 $ & $ -0.16 $ & $<$ & $ 0.30 $ & $ 0.10 $ & $ 0.15 $ \\
CS 22963-0004 & $ -3.03 $ & $ -2.57 $ & $<$ & $ -2.54 $ & $ -0.25 $ &  & - & - & $<$ & $ 0.61 $ & $ 0.13 $ & $ 0.37 $ \\
HE 0132-2439 & $ -2.52 $ & $ -2.64 $ &  & - & - & $<$ & $ -2.77 $ & $ -0.16 $ & $<$ & $ 0.01 $ & $ 0.32 $ & $ -0.15 $ \\
HE 0344-0243 & $ -3.00 $ & $ -2.69 $ & $<$ & $ -3.05 $ & $ -0.15 $ &  & - & - & $<$ & $ 0.28 $ & $ 0.33 $ & $ 0.12 $ \\
HE 0926-0546 & $ -3.55 $ & $ -3.06 $ & $<$ & $ -2.80 $ & $ -0.16 $ & $<$ & $ -2.91 $ & $ -0.15 $ & $<$ & $ 0.87 $ & $ 0.24 $ & $ 0.72 $ \\
\enddata
\tablecomments{Measured [Fe/H], [Ca/H], and [K/H] abundances in the sample stars. Measurements or upper limits are shown for each of the K I lines. $\Delta_{\rm NLTE}$ indicates the NLTE correction computed by interpolating the grid of Reggiani et al. (2019).}
\end{deluxetable*}

\subsection{Abundance ratios of other elements [X/Fe]}

\begin{figure*}[htbp]
    \centering
    \includegraphics[width=0.95\linewidth]{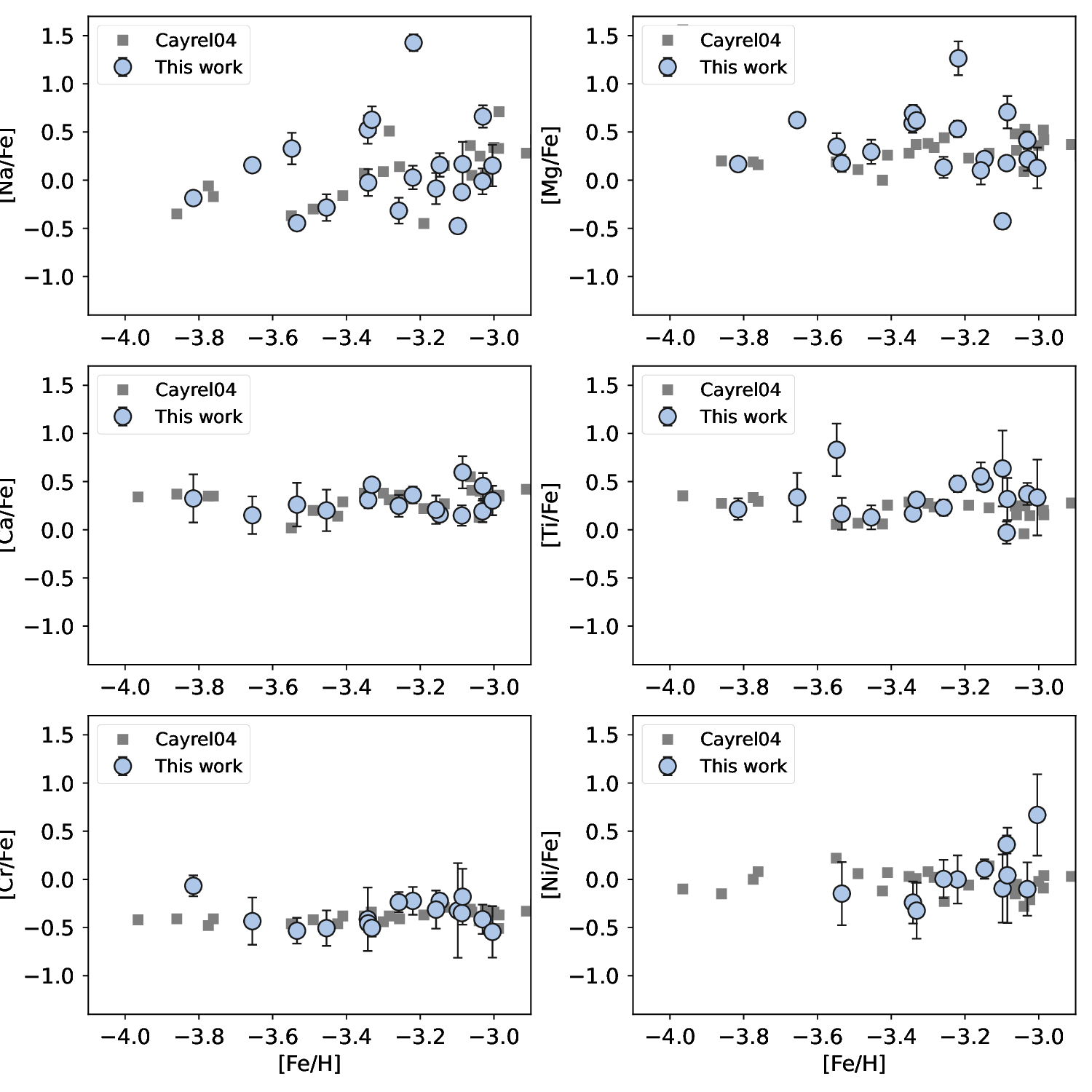}
    \caption{Abundances of other elements measured in this work. The gray squares show the results from \citet{Cayrel_2004A&A...416.1117C}.}
    \label{fig:xfe_feh}
\end{figure*}

Figure \ref{fig:xfe_feh} shows the abundance ratios of elements, other than K, as measured in this work. For comparison, the values obtained by \citet{Cayrel_2004A&A...416.1117C} are shown. 
The [X/Fe] trends with respect to [Fe/H], as obtained in this work, are in good agreement with those of \citet{Cayrel_2004A&A...416.1117C}. The Ni abundance ratios shown in the bottom-right panel exhibit a hint of a systematic offset at the lowest metallicities. This is likely due to the different \ion{Ni}{1} lines adopted in \citet{Cayrel_2004A&A...416.1117C}, for the same reason as discussed in Section \ref{sec:other_elem}.

\begin{deluxetable*}{lcccccccccccccc}\label{tab:abundance}
\tablecaption{Derived abundances of Fe, Na, Mg, Ca, Ti, Cr, and Ni.}
\tablehead{\colhead{Name} & \colhead{[Fe/H]} & \colhead{$\sigma_{[Fe/H]}$} & \colhead{[Na/Fe]} & \colhead{$\sigma_{[Na/Fe]}$} & \colhead{[Mg/Fe]} & \colhead{$\sigma_{[Mg/Fe]}$} & \colhead{[Ca/Fe]} & \colhead{$\sigma_{[Ca/Fe]}$} & \colhead{[Ti/Fe]} & \colhead{$\sigma_{[Ti/Fe]}$} & \colhead{[Cr/Fe]} & \colhead{$\sigma_{[Cr/Fe]}$} & \colhead{[Ni/Fe]} & \colhead{$\sigma_{[Ni/Fe]}$}\\\colhead{-} & \colhead{(dex)} & \colhead{(dex)} & \colhead{(dex)} & \colhead{(dex)} & \colhead{(dex)} & \colhead{(dex)} & \colhead{(dex)} & \colhead{(dex)} & \colhead{(dex)} & \colhead{(dex)} & \colhead{(dex)} & \colhead{(dex)} & \colhead{(dex)} & \colhead{(dex)}}
\startdata
BS 16076-0006 & $ -3.34 $ & $ 0.03 $ & $ 0.53 $ & $ 0.15 $ & $ 0.59 $ & $ 0.10 $ & - & - & - & - & $ -0.41 $ & $ 0.33 $ & - & - \\
BS 16550-087 & $ -3.34 $ & $ 0.01 $ & $ -0.03 $ & $ 0.14 $ & $ 0.69 $ & $ 0.09 $ & $ 0.31 $ & $ 0.09 $ & $ 0.17 $ & $ 0.07 $ & $ -0.46 $ & $ 0.09 $ & $ -0.24 $ & $ 0.22 $ \\
BS 16929-0005 & $ -3.22 $ & $ 0.01 $ & $ 0.03 $ & $ 0.12 $ & $ 0.53 $ & $ 0.09 $ & $ 0.36 $ & $ 0.09 $ & $ 0.48 $ & $ 0.09 $ & $ -0.22 $ & $ 0.14 $ & $ -0.00 $ & $ 0.25 $ \\
HE 1012-1540 & $ -3.22 $ & $ 0.03 $ & $ 1.43 $ & $ 0.09 $ & $ 1.26 $ & $ 0.17 $ & - & - & - & - & - & - & - & - \\
SDSSJ134338.67+484426.6 & $ -3.10 $ & $ 0.01 $ & $ -0.48 $ & $ 0.08 $ & $ -0.43 $ & $ 0.08 $ & - & - & $ 0.64 $ & $ 0.39 $ & $ -0.32 $ & $ 0.49 $ & $ -0.10 $ & $ 0.35 $ \\
BD+44 493 & $ -3.65 $ & $ 0.01 $ & $ 0.16 $ & $ 0.06 $ & $ 0.62 $ & $ 0.05 $ & $ 0.15 $ & $ 0.20 $ & $ 0.34 $ & $ 0.25 $ & $ -0.43 $ & $ 0.24 $ & - & - \\
CD-30 298 & $ -3.09 $ & $ 0.01 $ & $ -0.12 $ & $ 0.07 $ & $ 0.18 $ & $ 0.06 $ & $ 0.15 $ & $ 0.11 $ & $ -0.03 $ & $ 0.12 $ & $ -0.35 $ & $ 0.06 $ & $ 0.36 $ & $ 0.09 $ \\
CD-38 245 & $ -3.81 $ & $ 0.01 $ & $ -0.19 $ & $ 0.07 $ & $ 0.17 $ & $ 0.07 $ & $ 0.33 $ & $ 0.25 $ & $ 0.21 $ & $ 0.11 $ & $ -0.07 $ & $ 0.11 $ & - & - \\
CS 22189-0009 & $ -3.26 $ & $ 0.01 $ & $ -0.32 $ & $ 0.13 $ & $ 0.13 $ & $ 0.11 $ & $ 0.25 $ & $ 0.11 $ & $ 0.23 $ & $ 0.08 $ & $ -0.24 $ & $ 0.10 $ & $ 0.00 $ & $ 0.20 $ \\
CS 22172-0002 & $ -3.53 $ & $ 0.01 $ & $ -0.45 $ & $ 0.07 $ & $ 0.17 $ & $ 0.09 $ & $ 0.26 $ & $ 0.23 $ & $ 0.17 $ & $ 0.17 $ & $ -0.53 $ & $ 0.13 $ & $ -0.15 $ & $ 0.33 $ \\
CS 22942-0002 & $ -3.03 $ & $ 0.01 $ & $ -0.01 $ & $ 0.13 $ & $ 0.41 $ & $ 0.09 $ & $ 0.19 $ & $ 0.11 $ & $ 0.37 $ & $ 0.11 $ & $ -0.41 $ & $ 0.15 $ & $ -0.10 $ & $ 0.28 $ \\
CS 22949-0048 & $ -3.15 $ & $ 0.01 $ & $ 0.16 $ & $ 0.12 $ & $ 0.22 $ & $ 0.06 $ & $ 0.16 $ & $ 0.07 $ & $ 0.48 $ & $ 0.04 $ & $ -0.22 $ & $ 0.08 $ & $ 0.11 $ & $ 0.10 $ \\
CS 22950-0046 & $ -3.45 $ & $ 0.01 $ & $ -0.28 $ & $ 0.14 $ & $ 0.29 $ & $ 0.12 $ & $ 0.20 $ & $ 0.21 $ & $ 0.13 $ & $ 0.12 $ & $ -0.51 $ & $ 0.18 $ & - & - \\
CS 22963-0004 & $ -3.03 $ & $ 0.02 $ & $ 0.66 $ & $ 0.12 $ & $ 0.22 $ & $ 0.12 $ & $ 0.46 $ & $ 0.13 $ & - & - & - & - & - & - \\
CS 30339-0073 & $ -3.16 $ & $ 0.02 $ & $ -0.09 $ & $ 0.16 $ & $ 0.10 $ & $ 0.15 $ & $ 0.21 $ & $ 0.15 $ & $ 0.56 $ & $ 0.14 $ & $ -0.31 $ & $ 0.20 $ & - & - \\
HE 0130-1749 & $ -3.08 $ & $ 0.02 $ & $ 0.17 $ & $ 0.23 $ & $ 0.71 $ & $ 0.17 $ & $ 0.60 $ & $ 0.17 $ & $ 0.32 $ & $ 0.22 $ & $ -0.18 $ & $ 0.29 $ & $ 0.04 $ & $ 0.49 $ \\
HE 0132-2439 & $ -2.52 $ & $ 0.19 $ & $ -1.20 $ & $ 0.37 $ & - & - & - & - & - & - & - & - & - & - \\
HE 0344-0243 & $ -3.00 $ & $ 0.03 $ & $ 0.15 $ & $ 0.21 $ & $ 0.13 $ & $ 0.21 $ & $ 0.30 $ & $ 0.15 $ & $ 0.34 $ & $ 0.39 $ & $ -0.55 $ & $ 0.27 $ & $ 0.67 $ & $ 0.42 $ \\
HE 0926-0546 & $ -3.55 $ & $ 0.02 $ & $ 0.33 $ & $ 0.16 $ & $ 0.35 $ & $ 0.14 $ & - & - & $ 0.83 $ & $ 0.27 $ & - & - & - & - \\
SMSS J085924.06-120104.9 & $ -3.33 $ & $ 0.01 $ & $ 0.63 $ & $ 0.14 $ & $ 0.62 $ & $ 0.08 $ & $ 0.47 $ & $ 0.07 $ & $ 0.31 $ & $ 0.05 $ & $ -0.50 $ & $ 0.09 $ & $ -0.32 $ & $ 0.29 $ \\
\enddata

\tablecomments{Elemental abundance ratios of Fe, Na, Mg, Ca, Ti, Cr, and Ni for the sample stars. Statistical uncertainties are also shown.}\end{deluxetable*}

\subsection{Abundance patterns}

\begin{figure*}
    \centering
    \includegraphics[width=0.95\linewidth]{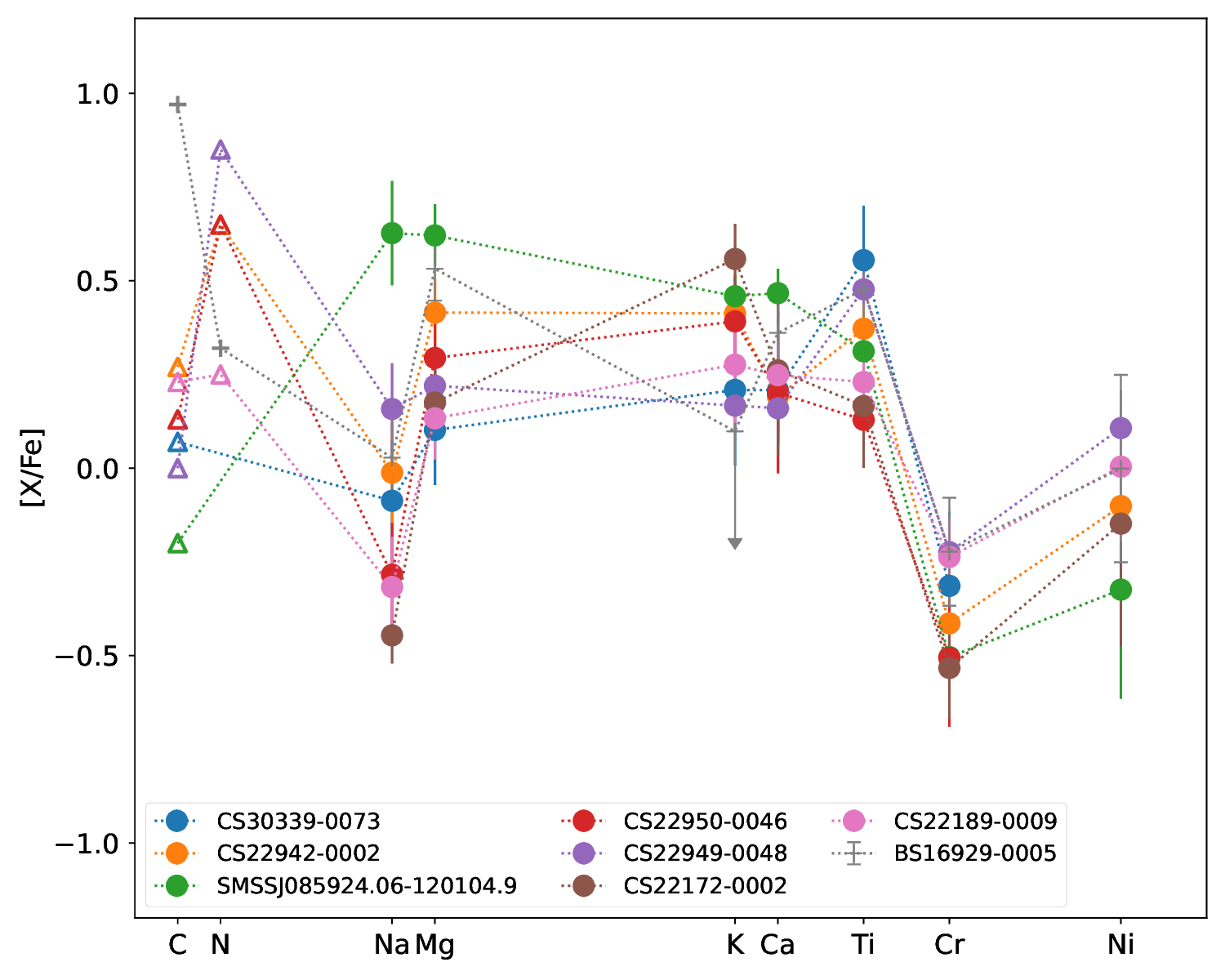}
    \caption{The abundance patterns for stars with estimated K abundances. The gray symbols indicate the stars whose upper limits were found to be lower than typical [K/Fe] values. The abundances of C and N are taken from literature (see text for details).}
    \label{fig:xfe}
\end{figure*}

Figure \ref{fig:xfe} presents the abundance patterns of stars with estimated [K/Fe]$_{\rm NLTE}$ ratios for Na, Mg, K, Ca, Ti, Cr, and Ni, as determined in this study. For reference,  the abundances of C and N are adopted from \citet{Jacobson_2015} for SMSS J085924.06-120104.9, and from \citet{Roderer_2014AJ....147..136R} for the other stars. None of those stars exhibited enhancements in Sr, Y, or Ba above solar values \citep{Roderer_2014AJ....147..136R, Jacobson_2015}, and thus these elements are not shown in the figure. 

As discussed in Section \ref{sec:kfe_kca}, the scatters in the [K/Fe]$_{\rm NLTE}$ and [K/Ca]$_{\rm NLTE}$ ratios are relatively low, estimated at $\sim$\qty{0.1} dex. In contrast, the [Na/Fe] ratios show a larger scatter of 0.34 dex. When Mg is used as a reference element, the scatter in [Na/Mg] is as large as 1.45 dex. 
The scatter is partly caused by NLTE effects on Na and Mn abundances \citep[][and reference therein]{Koutsouridou2025A&A...699A..32K}. After correcting for the NLTE effects using the \texttt{NiLITE} code provided by \citet{Koutsouridou2025A&A...699A..32K} using the grid of \citet{Lind2022A&A...665A..33L}, the scatter 
in [Na/Mg]$_{\rm NLTE}$ ratios is reduced to 0.74 dex. However, the scatter in the [Na/Mg]$_{\rm NLTE}$ ratios remains large compared to the scatter in [K/Ca]$_{\rm NLTE}$ ratios. 

SMSS J085924.06-120104.9 displays a particularly high [Na/Fe] ratio of 0.62 dex and [Na/Fe]$_{\rm NLTE} = 0.20$ dex, which largely contributes to the observed scatter. 
The star also exhibited the lowest [C/Fe] ratio among those with a detected \ion{K}{1} line. The low [C/Fe] is likely due to stellar internal mixing, as the evolutionary correction of the carbon abundance proposed by \citet{Placco2014ApJ...797...21P} can be as large as 0.4 dex for a star with $\log g = 1.65$ and [Fe/H]~$=-3.3$. In contrast to carbon, the observed Na abundance likely reflects the star's initial composition, since Na is not expected to be affected by internal mixing processes.

For BS 16929-0005 (gray symbols in Figure \ref{fig:xfe}), we obtained a relatively low upper limit for the K abundance. This star has previously been reported to be carbon enhanced, with [C/Fe]$=1.08$ by \citet{Aoki_2007ApJ...655..492A}, and a similar value of  [C/Fe]$=0.97$ by \citet{Lai_2008ApJ...681.1524L}.

The implications of these findings are discussed further in Section \ref{sec:namg}.

\subsection{Correlation with other odd-Z elements}

\begin{figure*}
    \centering
    \includegraphics[width=0.95\linewidth]{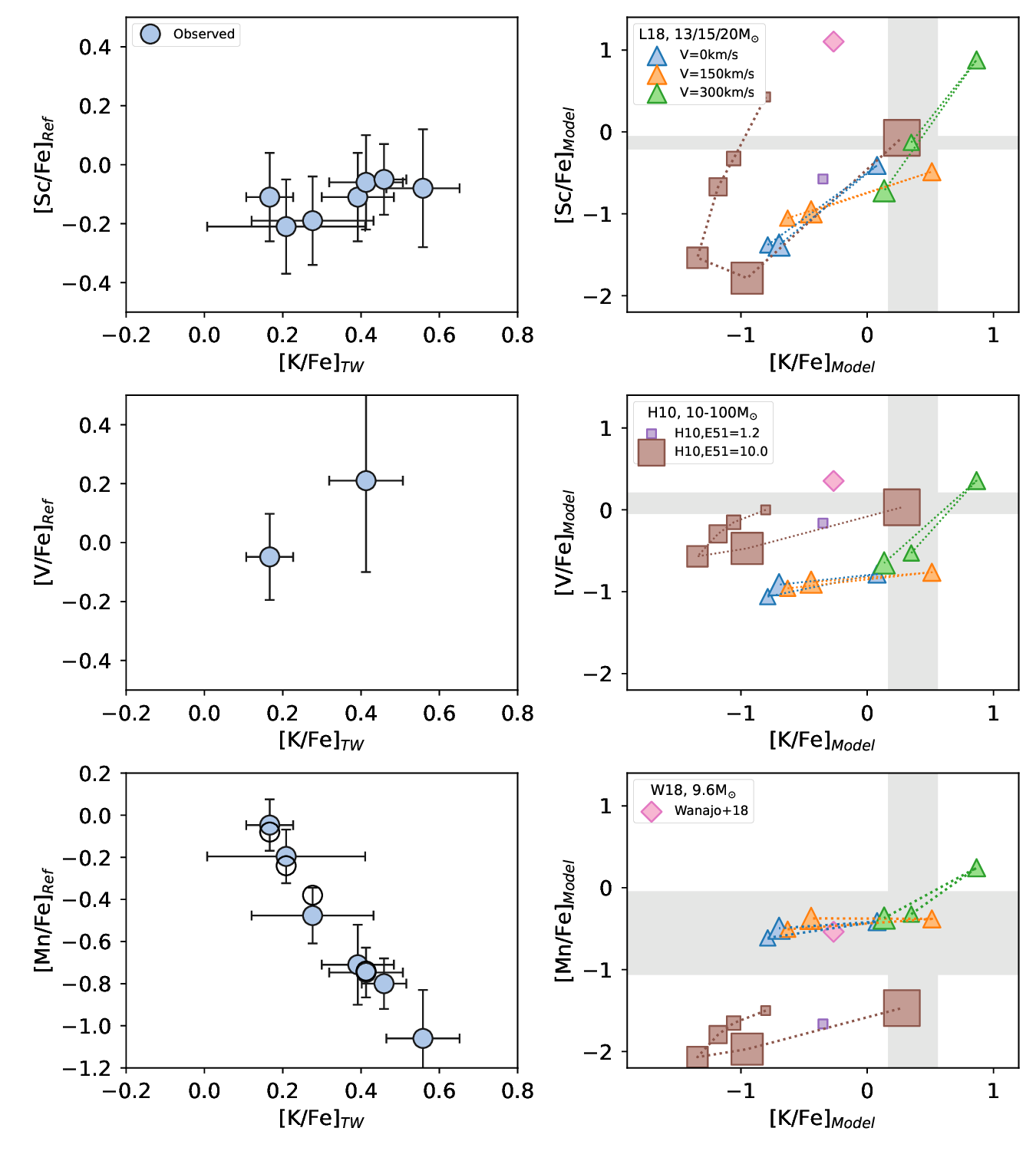}
    \caption{Left: The [Sc/Fe], [V/Fe], and [Mn/Fe] ratios adopted from the literature plotted against [K/Fe]$_{\rm NLTE}$ derived in this study. In the bottom-left panel for the [Mn/Fe] ratios plotted against [K/Fe]$_{\rm NLTE}$, we also show [Mn/Fe] ratios obtained using \ion{Mn}{2} lines only by \citet{Roderer_2014AJ....147..136R} for four of the sample stars in open circles.   Right: the predicted theoretical yields of odd-Z elements. Triangles are the rotating massive star yield model by \citet{Limongi_2018ApJS..237...13L}. The models with initial rotational velocities of \qty{0}{\kilo\meter\per\second}, \qty{150}{\kilo\meter\per\second}, and \qty{300}{\kilo\meter\per\second} are shown by blue, orange, and green triangles, respectively. For each rotational velocity, the models with three different progenitor masses, $M=13, 15$ and $20M_\odot$, are shown, where the larger symbol corresponds to a more massive model. Squares indicate the non-rotating supernova yield models with explosion energies of $E_{51}=1.2$ and $10$ from \citet{Heger_2010ApJ...724..341H}. The symbol size indicates the progenitor mass from 10 to 100$M_\odot$. The diamond shows the yields from the two-dimensional core-collapse supernova simulation of a 9.6$M_\odot$ star by \citet{Wanajo_2018ApJ...852...40W}. The observed range of each of the abundance ratios is shown by a vertical or horizontal gray band.}
    \label{fig:k_scvmn}
\end{figure*}

To explore potential correlations between the estimated potassium abundances (K) and those of other odd-Z elements, the left panels of Figure \ref{fig:k_scvmn} present the abundance ratios [Sc/Fe], [V/Fe], and [Mn/Fe], adopted from \citet{Roderer_2014AJ....147..136R}, \citet{Jacobson_2015} and \citet{Lai_2008ApJ...681.1524L}, respectively, plotted against the [K/Fe]$_{\rm NLTE}$ ratios derived in this study.

The [Sc/Fe] and [V/Fe] ratios (top-left and middle-left panels) appear to be relatively consistent among the stars with measured K abundances. In contrast, the [Mn/Fe] ratio (bottom-left panel) exhibits a possible anti-correlation with [K/Fe]$_{\rm{NLTE}}$. However, the weighted correlation coefficient of $-0.99$ is not statistically significant, as indicated by a p-value of 0.59, calculated from 1000 bootstrap resamplings of the abundance data. Neither the [K/Fe] nor the [Mn/Fe] ratios show correlation with the values of $T_{\rm eff}$ or $\log g$, and thus the anticorrelation is unlikely to be caused by NLTE corrections on Mn abundances of stars with different stellar parameters. 
As a reference, the bottom-left panel of Figure \ref{fig:k_scvmn} also shows the [Mn/Fe] ratios derived using \ion{Mn}{2} lines by \citet{Roderer_2014AJ....147..136R}, which are known to be less affected by the NLTE effects \citep{Bergemann2019A&A...631A..80B}, for four sample stars. The anticorrelation remains unchanged when using only those four stars. 

A comparison between these observational results and theoretical predictions from stellar and supernova yield models is presented in Section \ref{sec:discussion}.

\section{Discussion}\label{sec:discussion}

The K abundances estimated for EMP stars, including stars with [Fe/H]$<-3.5$, provide important insights into the nucleosynthetic yields from individual massive stars \citep{Tominaga_2014ApJ...785...98T,Heger_2010ApJ...724..341H,Nomoto_2013ARA&A..51..457N}. 
Our new analysis of high-resolution spectra confirmed the previously reported supersolar values of the [K/Fe] ratios in the lowest [Fe/H] regimes \citep{Cayrel_2004A&A...416.1117C,Andrievsky_2010A&A...509A..88A,Takeda_2009PASJ...61..563T,Reggiani_2019A&A...627A.177R}. Almost all of the stars studied in this work show enhancements not greater than [K/Fe]$_{\rm NLTE}=0.7$ dex. We also obtained a constraint on the scatter in the [K/Fe] and [K/Ca] ratios for stars with detected K lines.

\subsection{Scatter in [Na/Mg] ratios}\label{sec:namg}

Among the seven stars with measured K abundances, we found a significant scatter in the [Na/Mg] ratios. 
This scatter is unlikely to be caused by internal mixing processes, as 
neither Na nor Mg is expected to be altered at the stellar surface during the evolutionary stages of these stars. Therefore, the observed variation likely reflects inhomogeneities in the metal-enriched gas from which the stars formed.
Sodium and magnesium are synthesized during hydrostatic burning in massive stars, and their relative abundances are sensitive to the progenitor star's mass. In addition, Na production is influenced by the progenitor's metallicity and rotational velocities \citep[e.g.,][]{Nomoto_2013ARA&A..51..457N}.
Assuming that the abundance patterns observed in each star purely reflect the yield of 
a single supernova of a first-generation (Population III) star with the mass of 10-30 $M_\odot$, 
the scatter in [Na/Mg] ratios could be attributed to variations in progenitor properties such as mass, metallicity, and rotation. In contrast, the apparent co-production of K and Ca, evidenced by the lack of scatter in [K/Ca] ratios, suggests that the yields of these elements are less sensitive to such progenitor characteristics.

\subsection{Astrophysical origins of K}\label{sec:origin_K}

Regarding the origin of K in the extremely metal-poor regime, two potential production sites were proposed: (1) the final evolutionary stages of massive stars, and (2) explosive nucleosynthesis during core-collapse supernovae.  In the following subsections, we examine these scenarios in light of our abundance estimates.

\subsubsection{The final stages of massive star evolution} 

Hydrostatic oxygen burning in massive stars, which takes place during the final evolutionary stages of such stars, has been proposed as a potential site for K production \citep{Woosley_1995ApJS..101..181W, Woosley2002RvMP...74.1015W}. In particular, nucleosynthesis processes unique to rotating massive stars are predicted to enhance the yields of odd-Z elements, including K \citep{Limongi_2018ApJS..237...13L, Roberti_2024ApJS..270...28R}. Chemical evolution models that incorporate yields from rotating massive stars 
are able to reproduce the observed [K/Fe] ratios, 
at least at low [Fe/H] regime \citep{Prantzos_2018MNRAS.476.3432P,Reggiani_2019A&A...627A.177R}.

To examine the scenario in which oxygen burning during the evolution of rotating massive stars serves as a dominant source of K in the observed EMP stars, Figure \ref{fig:k_ca_yields} presents the [K/Ca] ratios predicted by the one-dimensional stellar evolution and core-collapse supernova yield models of \citet{Limongi_2018ApJS..237...13L}. These models incorporate rotational mixing within the interiors of massive stars with initial masses ranging from 13 to $120~M_\odot$. 
Different symbol sizes represent models with progenitor masses of 13, 15, 20, and 25~$M_\odot$ at [Fe/H]~$=-3.0$, while different colors indicate initial rotational velocities of $V_{\rm rot}=$ \qty{0}, \qty{150}, \qty{300}{\kilo\meter\per\second}.
The models assume mixing below the base of the oxygen-burning shell and the ejection of \num{0.07}~$M_\odot$ of $^{56}$Ni ("set R"). Stars with initial masses greater than $25M_\odot$ are assumed to undergo a direct collapse into black holes, thereby not contributing to chemical enrichment.  
The observed $1\sigma$ and $2\sigma$ ranges of the [K/Ca] ratios derived in this study are indicated by dark and light gray shared regions, respectively.

 Among these models considered, the one with a progenitor mass of 15~$M_\odot$ and a rotational velocity of $V_{\rm rot}=$\qty{300}{\kilo\meter\per\second} (green triangles) shows the best agreement with the observed [K/Ca] ratios. The model with the same progenitor mass but a lower rotational velocity of  $V_{\rm rot}=$\qty{150}{\kilo\meter\per\second} (orange triangles) is reasonably consistent with the observations within the 2$\sigma$ range. In contrast, models with lower ($M=13M_\odot$) and higher ($M\geq 20M_\odot$) progenitor masses predict significantly lower [K/Ca] ratios, by more than $0.5$ dex. Furthermore, non-rotating models (blue triangles) systematically yield lower [K/Ca] ratios across all considered masses by $\gtrsim 0.5$ dex.

The right panels of Figure \ref{fig:k_scvmn} show the abundance ratios of [Sc/Fe], [V/Fe], and [Mn/Fe] plotted against [K/Fe]. These predictions shown by triangles are based on the rotating massive star yield models of \citet{Limongi_2018ApJS..237...13L}. As in Figure \ref{fig:k_ca_yields}, different colors indicate models with different initial rotational velocities ($V_{\rm rot}$).
Similar to the trends seen in the [K/Ca] ratios, the model with a progenitor mass of 15~$M_\odot$ and $V_{\rm rot} = $\qty{300}{\kilo\meter\per\second} (green triangles) predicts the highest abundances for K, Sc, V, and Mn.
However, in contrast to the [K/Ca] case, the model with $M = 13~M_\odot$ and $V_{\rm rot} = $\qty{300}{\kilo\meter\per\second} best reproduces not only the [K/Fe] ratios but also the ratios of [Sc/Fe] and [Mn/Fe].
The 15~$M_\odot$ model with $V_{\rm rot} = $\qty{300}{\kilo\meter\per\second} tends to overproduce all of these abundance ratios relative to the fixed amount of Fe (produced as $^{56}$Ni) assumed in this model set.
Across all rotational velocities, the models generally predict either no correlation or a weak positive correlation between [K/Fe] and [Mn/Fe], which is inconsistent with the tentative anti-correlation as suggested by the lower-left panel of this figure.

\citet{Roberti_2024ApJS..270...28R} attributed the enhanced synthesis of oxygen-burning products, including K, in metal-free $M~=15_\odot$ rotating star yield models to the mixing between the convective O shell and the products of C-shell burning. 
Their calculations include stellar evolution and explosion yields for massive stars (15 and 25~$M_\odot$) across a wide range of initial rotational velocities (\qty{0} - \qty{800}{\kilo\meter\per\second}) and metallicities ([Fe/H] = $-\infty$, $-5$, $-4$).  
The enhancement of oxygen-burning products is found to occur regardless of the progenitor's metallicity. Under the proposed scenario, which involves the C-O shell merger, the oxygen-burning products are less likely to undergo further processing via explosive nucleosynthesis during the supernova and are instead ejected \citep{Roberti_2024ApJS..270...28R}. 
This scenario may also explain the observed variation in the Na/Mg ratios among the stars analyzed in this study, which could reflect differing degrees of these mixing processes during the final phase of stellar evolution \citep{Roberti_2024ApJS..270...28R}.
Interactions between oxygen and carbon shells in massive stars have also been suggested to explain the observed abundance patterns of other odd-Z elements, such as phosphorus (P) and chlorine (Cl) \citep{Ritter_2018MNRAS.474L...1R}.

In summary, rotating massive stars could account for the observed abundances of K and other odd-Z elements in the sample stars if the ejecta from rotating massive stars with rotational velocities up to \qty{300}{\kilo\meter\per\second} significantly contributed to the enrichment of the interstellar medium. 
Further examination of this scenario through homogeneous analysis of CNO as well as neutron-capture elements would be necessary to confirm this scenario. 

\subsubsection{The innermost region of CCSNe}

Another possible site of K production has been suggested as explosive oxygen burning in the 
innermost regions of core-collapse supernovae. The nucleosynthesis yields of these mechanisms are difficult to predict because of the uncertainty in the mechanism of supernova explosion.

\begin{figure}
    \centering
    \includegraphics[width=0.95\linewidth]{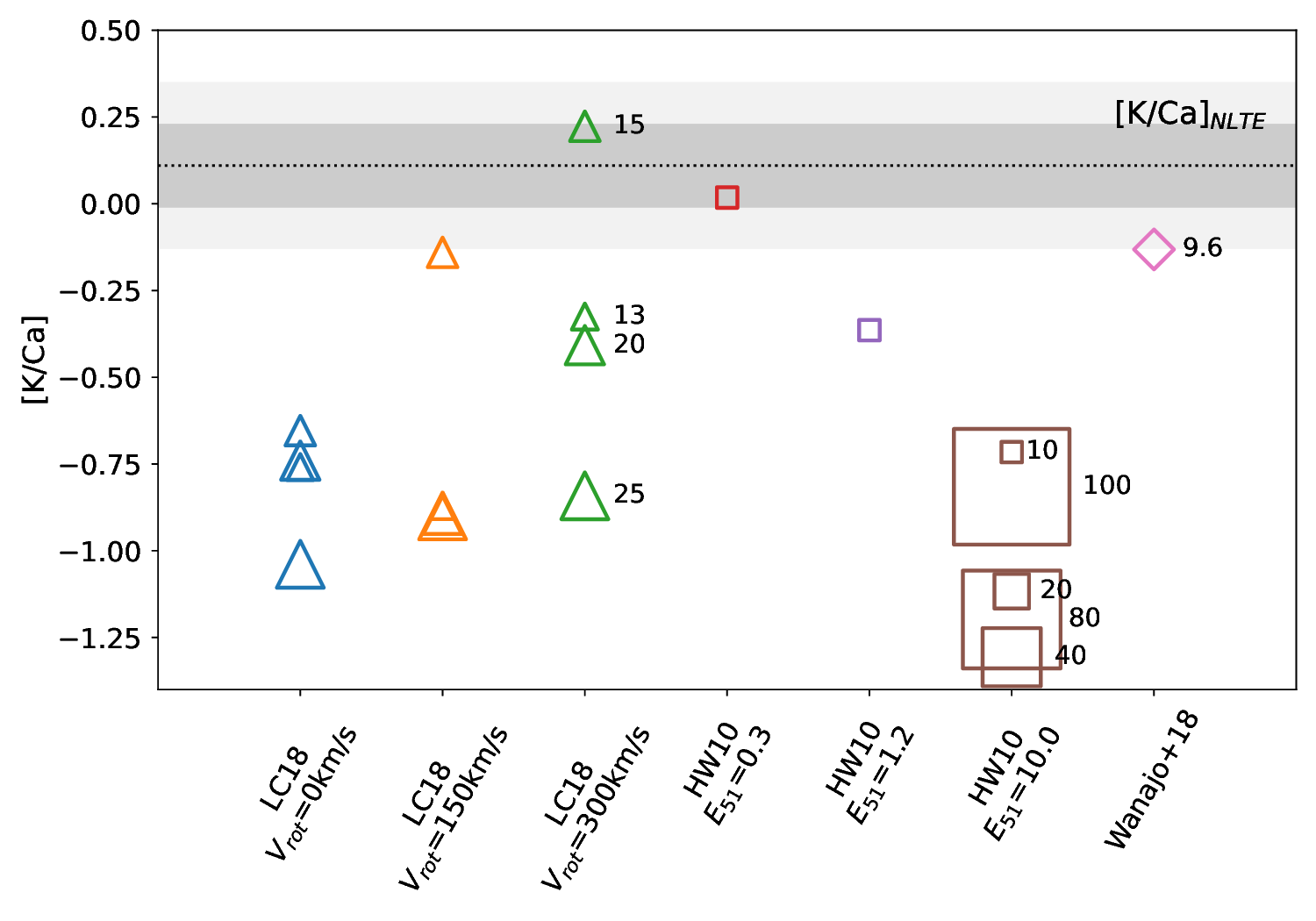}
    \caption{[K/Ca] abundance ratios predicted by core-collapse supernova yields in literature. Triangles correspond to the CCSN yields of rotating massive stars from \citet{Limongi_2018ApJS..237...13L}  with rotational velocity $V=$\qty{0} (blue), \qty{150} (orange), and \qty{300}{\kilo\meter\per\second} (green), respectively. The size of the symbols represents progenitor masses of 13, 15, 20, and 25$M_\odot$ and is indicated on the right-hand side of each symbol. Squares correspond to the CCSN yields of non-rotating zero-metallicity stars from \citet{Heger_2010ApJ...724..341H} with various supernova explosion energies. The size of the symbols represents progenitor masses of 10-100$M_\odot$ and is indicated on the right-hand side of each symbol. The diamond corresponds to the CCSN yields of a zero-metallicity $M=9.6M_\odot$ star based on a 2D simulation by \citet{Wanajo_2018ApJ...852...40W}. The mean of observed abundances (of 7 stars) and their 1$\sigma$ (dark gray) and 2$\sigma$ (light gray) are indicated by gray horizontal bands.}
    \label{fig:k_ca_yields}
\end{figure}

The square symbols in Figure \ref{fig:k_ca_yields} show the predicted CCSN yields of zero-metallicity nonrotating massive stars from \citet{Heger_2010ApJ...724..341H}, which includes $\nu$-process with assumed temperature and fluxes of neutrinos. We plotted the yields with an assumed mass cut at the base of the oxygen shell ("S4" model) with no mixing of ejecta. The models with extremely low yield of Ca $<10^{-5}M_\odot$ are excluded from this plot. The different colors correspond to the differences in the explosion energy, $E_51=0.3$, $1.2$, and $10.0$. The size of the symbols represents progenitor stellar masses $M=10, 20, 30, 40, 80$, and $100M_\odot$, where larger symbols correspond to higher stellar masses.

The model with a progenitor mass of 10$M_\odot$ and explosion energy $E_{51}=0.3$ predicts 
a K yield on the order of $\sim 10^{-6}M_\odot$, which overlaps within 1$\sigma$ of observed abundances. 
The models with the same stellar mass and higher explosion energies predict the K yield of $10^{-6}-10^{-5}M_\odot$, while the larger Ca yield in these models lead to [K/Ca] ratios much lower than the observed values.

Figure \ref{fig:k_scvmn} shows the correlation between the predicted [K/Fe] ratios and other odd-Z elements. 
The model with a progenitor mass of 10$M_\odot$ and explosion energy $E_{51}=0.3$ predicts extremely small Fe yields, and therefore incompatible with any of the abundance ratios shown in this figure. 
Instead, the model with high explosion energy $E_{51}=10.0$ and a progenitor mass of $100M_\odot$ predicts [Sc/Fe], [V/Fe], and [K/Fe] ratios that are compatible with the data. All models underestimate the [Mn/Fe] ratios.

It has long been recognized that the discrepancy between observed abundances and theoretical yield models based on one-dimensional core-collapse supernova simulations is due to the uncertainty in the elemental yields in the innermost region of the exploding star, which depends on the physical mechanism relevant to interactions between neutrinos and nucleons \citep{Frohlich_2006ApJ...637..415F,Wanajo_2018ApJ...852...40W}.  \citet{Yoshida_2008ApJ...672.1043Y} also suggests that the $\nu-$process 
 in core-collapse supernovae of Population III stars produces observed abundances of odd-Z elements such as K, Sc, V, and Mn
\citep{Yoshida_2008ApJ...672.1043Y, Kobayashi_2011}. 
With the $\nu$-process, the predicted values of [Mn/Fe] are enhanced to reconcile with the observed abundance ratios of EMP stars.

To reliably predict the yields of K in the innermost regions of core-collapse supernovae, multidimensional hydrodynamical simulations are necessary.  \citet{Wanajo_2018ApJ...852...40W} carried out nucleosynthesis calculations in the innermost, neutrino-heated ejecta from several 2-dimensional, self-consistently exploding models of core-collapse supernovae, from which we adopt the model yields of the zero-metallicity star with an initial mass of $9.6 M_\odot$ ("z9.6" model). Potassium 
appears 
to be strongly 
dependent on $Y_e$, and thus can be a tracer to test 
such multidimensional models. 
In their simulation, K is synthesized under proton-rich ($Y_e>0.5$) conditions in the innermost ejecta, while Ca is predominantly produced at $Y_e \sim 0.5$. Under such conditions, nucleosynthesis does not depend on the composition prior to the explosion, as the interaction between neutrinos and nucleons redetermines the electron fraction $Y_e$. This scenario of nucleosynthesis in neutrino-processed ejecta is also supported by recent X-ray measurements of supernova remnants \citep{Sato_2023ApJ...954..112S}.

The predicted abundance ratio [K/Ca]
is shown in Fig. \ref{fig:k_ca_yields} as a diamond symbol. The K and Ca yields from this simulation are $M=3.37\times 10^{-6} M_\odot$ and $M=9.58\times 10^{-5} M_\odot$, both of which are compatible with the observed ratios [K/H] and [Ca/H] if the ejecta are mixed with hydrogen of $10^3-10^4M_\odot$, although the [K/Ca] ratio is smaller than the observed values. This implies that the innermost ejecta of typical core-collapse supernovae are dominated by more proton-rich ejecta than that adopted in the present model. The result of the small K abundance scatter, despite a large scatter in the lighter elements, is also in line with this prediction from the K production in neutrino-processed ejecta,
where the yields are mainly determined by the physical conditions in the innermost ejecta of core-collapse supernovae with little dependence on progenitor mass or metallicity. 
The predicted abundance ratios for other odd Z elements are shown in the right panels of Figure \ref{fig:k_scvmn}. The model underproduces [K/Fe] ratios, while the [Mn/Fe] ratio is within the observed range.

Overall, none of the yield models of K synthesis in core-collapse supernovae examined in this section simultaneously reproduce the observed abundance ratios of K, Sc, V, and Mn. Furthermore, the $\nu$-process suggested to be the site of some of the odd-Z elements generally predicts a positive correlation between K and Mn abundance ratios, which is incompatible with the hint of anti-correlation observed among stars with a detected \ion{K}{1} line. Multidimensional simulations for a wide range of progenitor masses and explosion energies, taking into account the $\nu$-process in a self-consistent manner, would be essential to obtain a consensus on the origin of K at the lowest metallicity regime.

\section{Conclusion}\label{sec:conclusion}

We revisited the abundances of Potassium (K) in extremely metal-poor stars based on a new homogeneous analysis of K abundances derived from high-resolution spectra obtained with Subaru/HDS. NLTE corrections from \citet{Reggiani_2019A&A...627A.177R} were applied to derive the NLTE abundances based on the \ion{K}{1} \qty{766} and \qty{769}{nm} resonance lines.  This study demonstrates that the correlation between K abundances and other elemental abundance ratios in EMP stars provides important constraints on the late stages of massive stellar evolution and the unknown mechanisms of supernova explosions, both of which are inaccessible to direct observation and are challenging to theoretically simulate from first principles.

Our main findings can be summarized below. 

\begin{itemize}
    \item Among the seven stars with [Fe/H]$<-3$ with measured K abundances, the scatter in the [K/Fe] and [K/Ca] ratios is as small as the typical measurement uncertainty. After the NLTE correction, the mean abundance is in good agreement with the previous analysis of the NLTE K abundance by \citet{Takeda_2009PASJ...61..563T} and \citet{Reggiani_2019A&A...627A.177R}. 
    \item The observed upper limits of the K abundance ratios are [K/Fe]$_{\rm NLTE}<0.8$ dex and [K/Ca]$_{\rm NLTE}<0.7$, respectively, for most of the sample stars. 
    \item Despite the uniformity of the [K/Fe] and [K/Ca] ratios, the [Na/Mg] ratios show a scatter of 0.7 dex, which could be caused by variations in progenitor mass, metallicity of the core-collapse supernovae, or the various degrees of mixing of different layers at the end of massive star evolution. This suggests that the nucleosynthesis site of K is likely independent of those other properties. 
   \item  An apparent anti-correlation between the [K/Fe] and [Mn/Fe] abundance ratios has been identified. If this trend is confirmed with a larger statistical sample and reduced systematic uncertainties, it would provide a valuable diagnostic of the physical conditions governing the nucleosynthesis of these elements in the early universe.
\end{itemize}

The uniform abundance ratios of [K/Ca] among the EMP stars might suggest a universal origin of K at the lowest [Fe/H] regime. The comparison of the observed [K/Ca] ratios with nucleosynthesis yield models of massive stars suggests that, depending on the progenitor masses, rotational velocities, or explosion energies, some of the hydrostatic burning or core-collapse supernovae of (non-) rotating massive stars could be the site of K synthesis together with Ca. However, none of the yield models simultaneously reproduces all of the observed abundances of odd-Z elements.

To obtain a robust conclusion on the nucleosynthesis sites of K, together with other odd Z elements, a tighter constraint on the K abundance upper limits and a statistical sample of EMP stars with detailed elemental abundances are both important. Updated theoretical insights from multidimensional effects in stellar evolution and core-collapse supernova explosions are also crucial to ultimately constraining the K synthesis sites, which may be related to yet unknown physical mechanisms of dying massive stars.

\begin{acknowledgments}
We are grateful to the anonymous referee for their helpful suggestions, which significantly improved the quality of this paper. We thank Y. Takeda for his advice on proposal preparation and abundance analysis. We thank S. Katsuda, L. Roberti, and A. Karakas for their valuable discussions and constructive suggestions, which have greatly improved this work. 
MNI would like to thank S. Hong for her valuable contributions to the data analysis and discussions. This work was supported by JSPS KAKENHI Grant Numbers JP25K01046, JP23KF0290, JP21H04499, and JP20H05855. This work is based on observations made with the Subaru Telescope, operated by the National Astronomical Observatory of Japan. We are grateful to the observatory staff for their support during the observations. 
This work has made use of the VALD database, operated at Uppsala University, the Institute of Astronomy RAS in Moscow, and the University of Vienna.
\end{acknowledgments}

%

\vspace{5mm}
\facilities{Subaru Telescope (HDS)}


\software{astropy \citep{astropy_2013ascl.soft04002G},  
     IRAF \citep{Tody_1993ASPC...52..173T}, 
     PyRAF \citep{PyRAF_2012ascl.soft07011S}, 
     iSpec \citep{Blanco-Cuaresma_2014A&A...569A.111B, Blanco-Cuaresma_2019MNRAS.486.2075B}, 
     Turbospectrum \citep{Plez_2012ascl.soft05004P}
          }




\bibliography{ref}{}
\bibliographystyle{aasjournalv7}



\end{document}